\documentclass[12pt]{iopart}

\newcommand{\dd}{\,\mathrm{d}}

\usepackage{bm}
\usepackage{iopams}
\usepackage{braket}
\usepackage{cite}
\usepackage{longtable}
\usepackage{caption}
\usepackage{multirow}
\usepackage{graphicx}
\usepackage{ntheorem}
\usepackage{stackrel}
\usepackage{subfigure}
\usepackage{hhline}
\usepackage{epstopdf}
\usepackage[fleqn]{cases}
\usepackage{stackrel}

\newtheorem{proposition}{Proposition}
\newtheorem{corollary}{Corollary}
\newtheorem{remark}{Remark}

\newtheorem{lemma}{Lemma}

\begin{document}
\title{Entanglement capacity of fermionic Gaussian states}


\author{Youyi Huang and Lu Wei}
\address{Department of Computer Science, Texas Tech University, Texas 79409, USA}
\ead{\{youhuang,luwei\}@ttu.edu}

\vspace{9pt}

\begin{indented}
\item[February 2023]
\end{indented}

\begin{abstract}
We study the capacity of entanglement as an alternative to entanglement entropies in estimating the degree of entanglement of quantum bipartite systems over fermionic Gaussian states. In particular, we derive the exact and asymptotic formulas of average capacity of two different cases -- with and without particle number constraints. For the later case, the obtained formulas generalize some partial results of average capacity in the literature. The key ingredient in deriving the results is a set of new tools for simplifying finite summations developed very recently in the study of entanglement entropy of fermionic Gaussian states.
\end{abstract}

\vspace{2pc}
\noindent{\it Keywords}: quantum entanglement, entanglement capacity, fermionic Gaussian states, random matrix theory, orthogonal polynomials, special functions

\maketitle

\section{Introduction}\label{sec:intr}
Entanglement is a fundamental feature of quantum mechanics and it is also the resource that enables quantum information processing as an emerging technology. The understanding of entanglement is crucial to a successful exploitation of advances of the quantum revolution. In the past decades, there has been considerable progress in estimating the degree of entanglement over different models of generic states, where one of the most extensively studied area is the entropy based estimations using, for example, von Neumann entropy~\cite{HLW06,Page93,Foong94,Ruiz95,VPO16,Wei17,Wei20,HWC21,SK2019,Wei20BHA,Wei20BH}, quantum purity~\cite{SK2019,Lubkin78,Sommers04,Giraud07,Osipov10,LW21}, and Tsallis entropy~\cite{MML02,Wei19T} as entanglement indicators. These results mainly focus on the statistical behavior of entanglement entropies over generic state models, such as the well-known Hilbert-Schmidt ensemble~\cite{Lubkin78, Page93,HLW06,MML02,Foong94,Ruiz95,Giraud07,VPO16,Wei17,Wei19T,Wei20,HWC21}, the Bures-Hall ensemble~\cite{Sommers04,Osipov10,Borot12,SK2019,Wei20BHA,Wei20BH,LW21}, and the fermionic Gaussian ensemble~\cite{BHK21,HW22,BHKRV22,HW23}. Besides entropies, there is a growing interest in understanding the capacity of entanglement as another entanglement quantifier. Similarly to entanglement entropy as an analogy to the thermal entropy, the entanglement capacity introduced in~\cite{YQ10} serves as an analogy to thermal heat capacity. It is also identified as a critical value to distinguish integrable systems from chaotic ones~\cite{Nandy}. In the literature, different properties of entanglement capacity have been numerically studied in~\cite{Nandy,ADKT23}. Moreover, exact formulas of the average capacity are recently obtained for the Hilbert-Schmidt ensemble~\cite{Boer19,OKUYAMA21,Wei23} and the Bures-Hall ensemble~\cite{Wei23}. For the fermionic Gaussian ensemble without particle number constraint, the average capacity of equal subsystem dimensions is derived in~\cite{BNP21,HW22}, whereas the corresponding exact formula in the general case of unequal dimensions remains open.

In this work, we compute the exact average entanglement capacity valid for any subsystem dimensions of fermionic Gaussian states for the cases of with and without particle number constraints. A key ingredient in obtaining the results is the set of tools for simplifying finite summations developed very recently~\cite{HW23} in the study of von Neumann entropy of the fermionic Gaussian ensemble. Our exact results also lead to the limiting values of average capacity when the subsystem dimensions approach infinity with a fixed dimension difference. Simulations are performed to numerically verify the derived results.

The rest of the paper is organized as follows. In~\sref{sec:formulation}, we first outline the problem formulation before presenting our main results of the exact mean capacity of fixed particle numbers and arbitrary particle numbers in proposition~\ref{prop1} and proposition~\ref{prop2}, respectively. The corresponding asymptotic capacity formulas are given in corollary~\ref{corollary1}. Proofs to the results are provided in~\sref{sec:2}. In~appendix A, we list summation representations of the integrals involved in the proofs. Summation identities utilized in the simplification are listed in appendix~B. The coefficients of some intermediate results appeared in the derivation are provided in~appendix~C.

\section{Problem formulation and main results}\label{sec:formulation}
\subsection{Problem formulation}
We first introduce the formulation that leads to the entanglement capacity of fermionic Gaussian states with and without particle number constraints as well as the corresponding statistical ensembles.

A system of $N$ fermionic degree of freedom can be formulated in terms of a set of fermionic creation and annihilation operators $\hat{a}_{i}$ and $\hat{a}_{i}^{\dag}$, $i=1,\dots, N$, which obey the canonical anti-commutation relation,
\begin{equation}\label{facr}
\{\hat{a}_{i},\hat{a}_{j}^{\dag}\}=\delta_{ij}\mathbb{I}, \qquad \{\hat{a}_{i},\hat{a}_{j}\}=0=\{\hat{a}_{i}^{\dag},\hat{a}_{j}^{\dag}\},
\end{equation}
where $\{\hat{A},\hat{B}\}=\hat{A}\hat{B}+\hat{B}\hat{A}$ denotes the anti-commutation relation and $\mathbb{I}$ is an identity operator. These fermionic modes can be equivalently described via the Majorana operators $\gamma_l$, $l=1,\dots, 2N$ , and
\begin{equation}
\hat{\gamma}_{2i-1}=\frac{\hat{a}_{i}^\dag+\hat{a}_{i}}{\sqrt{2}},\qquad \hat{\gamma}_{2i}=\imath\frac{\hat{a}_{i}^\dag+\hat{a}_{i}}{\sqrt{2}}
\end{equation}
with $\imath=\sqrt{-1}$ denoting the imaginary unit.
The Majorana operators also satisfy the anti-commutation relation
\begin{equation}\label{macr}
\{\hat{\gamma}_{l},\hat{\gamma}_{k}\}=\delta_{lk}\mathbb{I}.
\end{equation}
 By collecting the Majorana operators into a $2N$ dimensional operator-valued column vector ${\gamma}=(\hat{\gamma}_1,\dots,\hat{\gamma}_{2N})^{\dag}$,
a system of fermionic Gaussian state is then characterized by the density operator of the form~\cite{ST21,BHKRV22}
\begin{equation}\label{Fgdo}
\rho(\gamma)=\frac{\mathrm{e}^{-\gamma^\dag Q\gamma}}{\tr(\mathrm{e}^{-\gamma^\dag Q\gamma})},
\end{equation}
where the coefficient matrix $Q$ is a $2N\times 2N$ imaginary anti-symmetric matrix as the consequence of the anti-communication relation~(\ref{macr}).

\subsection*{-- Entanglement capacity over fermionic Gaussian states without particle number constraint}
There always exists an orthogonal matrix $M$ that diagnoses the coefficient matrix $Q$ by transforming ${\gamma}$ into another Majorana basis $\mu=(\hat{\mu}_1,\dots,\hat{\mu}_{2N})^\dag=M\gamma$.
A fermionic Gaussian state of arbitrary particle numbers is determined by the anti-symmetric covariance matrix~\cite{BHKRV22}
\begin{equation}
 J=-\imath\tanh (Q)=M^T J_0M,
\end{equation}
where $\tanh(x)$ denotes the hyperbolic tangent function~\cite{AS72}, the matrix $J_0$ takes the block diagonal form
\begin{equation}\label{eq}
J_{0}=\left(\begin{array}{ccc}
\tanh(\lambda_1)\mathbb{A} & \dots & 0 \\
\vdots & \ddots & \vdots \\
0 & \dots & \tanh(\lambda_N)\mathbb{A} \\
\end{array}\right),
\end{equation}
and
\begin{equation}\label{eq}
\mathbb{A}=\left(\begin{array}{cc}
0 & 1 \\
-1 & 0 \\
\end{array}\right).
\end{equation}
In the setting of the quantum bipartite model~\cite{BZ06}, the system of $N$ fermionic degree of freedoms can be decomposed into two subsystems $A$ and $B$ of dimension $m$ and $n$, respectively, with $m+n=N$. We assume $m\leq n$ without loss of generality. By restricting the matrix $J$ to the entries from subsystem $A$, the restricted covariance matrix $J_A$ is the $2m\times 2m$ left-upper block of $J$. The entanglement capacity can be represented via the real positive eigenvalues $x_i,~i=1,\dots,m$ of $\imath J_{A}$ as~\cite{Nandy,ADKT23,BNP21}
\begin{eqnarray}
C&=&\sum_{i=1}^{m}u(x_i)\label{eq:capacity}
\end{eqnarray}
with
\begin{eqnarray}\label{eq:ux1}
u(x)&=&\frac{1-x^{2}}{4}\ln^2\frac{1+x}{1-x}.
\end{eqnarray}
The resulting joint probability density of the eigenvalues $x_i,~i=1,\dots,m$ is proportional to \cite{BHK21}
\begin{equation}\label{eq:ensemble_ap}
\prod_{1\leq i<j\leq m}\left(x_{i}^2-x_{j}^2\right)^{2}\prod_{i=1}^{m}\left(1-x_{i}^2\right)^{n-m}, \qquad x_{i}\in[0,1],
\end{equation}
which is obtained by recursively applying the result in~\cite[proposition A.2]{KFI19}.

\subsection*{-- Entanglement capacity over fermionic Gaussian states with particle number constraint}
For a fermionic Gaussian state $\ket{F}$  with a fixed particle number $p$, it is more convenient to formulate it with the fermionic creation and annihilation operators, and the corresponding covariance matrix can be expressed as~\cite{BHKRV22,LRV20,LRV21}
\begin{equation}
H_{ij}=-\imath\bra{F}\hat{a}_i^{\dag}\hat{a}_j-\hat{a}_j\hat{a}_i^{\dag}\ket{F}.
\end{equation}
Using the anti-commutation relation~(\ref{facr}), the entries of the matrix $H$ then become
\begin{equation}\label{Fcov}
H_{ij}=-2\imath G_{ij}+\imath\delta_{ij},
\end{equation}
where $G_{ij}=\bra{F}\hat{a}_i^{\dag}\hat{a}_j\ket{F}$ denotes the entries of an $N\times N$ matrix $G$. There always exists a unitary transformation $U$ that diagonalizes $G$. In the resulting diagonal form, the first $p$ elements are equal to $1$ and the rest are $0$. Therefore, one can write
\begin{equation}
G=U_{N\times p}U_{N\times p}^\dag.
\end{equation}
Denoting $y_i$, $i=1,\dots,m$ the eigenvalues of the restricted matrix $G_A=U_{m\times p}U_{m\times p}^\dag$, the entanglement capacity can be represented as the function of $y_i$ as~\cite{ADKT23}
\begin{equation}~\label{eq:caf}
C=-\sum_{i=1}^m u(2y_i-1), \qquad y_{i}\in[0,1].
\end{equation}
The eigenvalue distribution of the random matrix $U_{m\times p}U_{m\times p}^\dag$ is the well-known Jacobi unitary ensemble~\cite{Mehta, Forrester}. Here, it is more convenient to use the  the eigenvalues of matrix $\imath H$. Denote $x_i$, $i=1,\dots,m$, as the eigenvalues of the $m\times m$ upper-left block of the matrix $\imath H$, the change of variables $x_i=2y_i-1$ in~(\ref{eq:caf}) leads to the entanglement capacity~(\ref{eq:capacity}) for the case of fixed particle number. The resulting joint probability density of the eigenvalues $x_i, i=1,\dots, m$, is proportional to~\cite{BP21}
\begin{equation}\label{eq:ensemble_fp}
\prod_{1\leq i<j\leq m}\left(x_{i}-x_{j}\right)^{2}\prod_{i=1}^{m}\left({1+x_i}\right)^{p-m}\left(1-x_{i}\right)^{n-p},\qquad x_{i}\in[-1,1].
\end{equation}

It has been introduced in~\cite{HW23} that the joint probability densities~(\ref{eq:ensemble_ap}) and~(\ref{eq:ensemble_fp}) can be compactly represented by a single joint density as
\begin{equation}\label{eq:fg-ensemble}
f_{\mathrm{FG}}(x) \propto \prod_{1\leq i<j\leq m}\left(x_{i}^\gamma-x_{j}^\gamma\right)^{2}\prod_{i=1}^{m}\left(1-x_{i}\right)^{a}\left(1+x_{i}\right)^{b}.
\end{equation}
The two considered scenarios of fermionic Gaussian states can now be conveniently identified by the above density~\eref{eq:fg-ensemble}, where
we have
\begin{equation}\label{eq:fgfp}
\gamma=1,~~~ a=n-p\geq 0,~~~ b=p-m\geq 0,~~~x\in[-1,1]
\end{equation}
for fermionic Gaussian states with an arbitrary number of particles, and
\begin{equation}\label{eq:fgap}
\gamma=2,~~~ a=b=n-m\geq 0,~~~x\in[0,1]
\end{equation}
for fermionic Gaussian states with a fixed number of particles. Note that computing the average capacity for the two cases will be performed separately below since the computation for an arbitrary $\gamma$ in~(\ref{eq:fg-ensemble}) appears difficult. We omit the normalization constants in~\eref{eq:fg-ensemble} as they will not be utilized in the calculation.

\subsection{Main results}
We now present our main results on the exact and asymptotic average capacity of the fermionic Gaussian states for the cases of fixed and arbitrary number of particles.
\begin{proposition}\label{prop1}
Denote the summation $\Phi _{c,d}$ as
\begin{equation}\label{eq:sumpsiab}
\Phi _{c,d}=\frac{c! }{(c+d)!}\sum _{k=1}^c \frac{(c+d-k)!}{ (c-k)!}\frac{1}{k^2}, \qquad c,d\in \mathbb{Z^+},
\end{equation}
and the function $F(a,b)$ as
\begin{eqnarray}
\fl F(a,b)&=&\alpha_0 \bigg(2 \Phi _{a+m,b}+2 \Phi _{m,a}+\psi _1(a+b+m+1)+\psi _1(a+m+1)+(\psi _0(a+m+1)\nonumber\\
\fl &&\!-\psi _0(a+b+m+1)){}^2-\psi _1(1)\bigg)+\alpha_1 \psi _0(a+m+1)+\alpha_2 \psi _0(a+1)+\alpha_3, \label{eq:prop1f}
\end{eqnarray}
where the coefficients  $\alpha_i$ are
\begin{eqnarray}
\alpha_0&=&\frac{m (a+m) (b+m) (a+b+m)}{(a+b+2 m-1)_3}\label{coef:c0}\\
\alpha_1&=&\frac{(a+b) (a+m-1) (a+m)}{(a+b+2 m-1)_2}\label{coef:c1}\\
\alpha_2&=&-\frac{a \left(a^2+a b+2 a m-a+2 b m-b+2 m^2-2 m\right)}{(a+b+2 m-1)_2}\label{coef:c2}\\
\alpha_3&=&\frac{m (a+m-1)}{(a+b+2 m-1)_2}-\frac{m}{2}\label{coef:c3}.
\end{eqnarray}
Then, for any subsystem dimensions $m\leq n$, the mean value of entanglement capacity~\eref{eq:capacity} of fermionic Gaussian states with a fixed particle number~\eref{eq:fgfp} is given by
\begin{equation}\label{eq:prop1}
\mathbb{E}\!\left[C\right]=F(p-m,n-p)+F(n-p,p-m).
\end{equation}
\end{proposition}

In proposition~\ref{prop1},
\begin{equation}\psi_{0}(x)=\frac{\dd\ln\Gamma(x)}{ \dd x}\end{equation}
and
\begin{equation}\psi_{1}(x)=\frac{\dd^2\ln\Gamma(x)}{ \dd^2 x}\end{equation} denote respectively the digamma and trigamma functions, and
\begin{equation}(a)_{n}=\frac{\Gamma(a+n)}{\Gamma(a)}\end{equation}
denotes the Pochhammer symbol.
The proof of proposition~\ref{prop1} can be found in~\sref{sec:2.1}. Note that the summation $\Phi _{c,d}$ in~\eref{eq:sumpsiab} does not  in general admit a closed-form representation for arbitrary $c$ and $d$. On the other hand, the sum $\Phi _{c,d}$ may be further simplified in some special cases as discussed in the following remark.

\begin{remark}\label{remark1}
Substituting $i \to k$, $m\to c$, $n \to c+d$ in the identity~\eref{eq:B12}, the summation $\Phi _{c,d}$ in~\eref{eq:sumpsiab} admits an alternative form
\begin{equation}\label{eq:unsimb1}
\sum_{k=1}^{c}\frac{\psi_0(k+d)}{k}+\mathrm{CF},
\end{equation}
where $\mathrm{CF}$ denotes the closed-form terms in the bracket of~\eref{eq:B12}. The sum in~\eref{eq:unsimb1} may not be summable into a closed-form expression and is referred to as an unsimplifiable basis~\cite{Wei17, Wei20, Wei20BH, HWC21, HW22, Wei23, HW23}. However, in the special cases of a given integer $d$, it permits closed-form representation as a result of the identity~\eref{eq:B3}. This corresponds to the case of fixed differences $a=n-p$, $b=p-m$, where the average capacity~\eref{eq:prop1} admits more explicit expressions. The cases $a=b=0, 1, 2$ are provided respectively in below as examples
\begin{eqnarray}
\fl \mathbb{E}\!\left[C\right]&=&-\frac{2 m^3}{(2m-1)(2m+1)}\left(\psi _1(m+1)-\frac{\pi ^2}{4}\right)-\frac{2 m^2-2 m+1}{2 m-1}\\
\fl \mathbb{E}\!\left[C\right]&=&-\frac{2 m (m+1) (m+2)}{(2 m+1) (2 m+3)}\left(\psi _1(m+1)-\frac{\pi ^2}{4}\right)-\frac{m (2 m (m+3)+5)}{(m+1) (2 m+3)}\\
\fl \mathbb{E}\!\left[C\right]&=&-\frac{2 m (m+2) (m+4)}{(2 m+3) (2 m+5)}\left(\psi _1(m+1)-\frac{\pi ^2}{4}\right)+\frac{4}{(m+1) (m+3)}\nonumber\\
\fl&&\times\left(\psi _0(m+1)-\psi _0(1)\right)-\frac{m \left(m^2+4 m+5\right) \left(4 m^3+30 m^2+72 m+57\right)}{(2 m+3) (2 m+5) (m+1)_3}.
\end{eqnarray}
\end{remark}

\begin{proposition}\label{prop2}
For any subsystem dimensions $m\leq n$, the mean value of entanglement capacity~\eref{eq:capacity} of fermionic Gaussian states with an arbitrary particle number~\eref{eq:fgap} is given by
\begin{eqnarray}\label{eq:prop2}
\fl \mathbb{E}\!\left[C\right]&=&\beta \left(\Phi _{2 m-1,n-m}+\Phi _{m+n-1,n-m}\right)+\frac{1}{4} \left(\Phi _{m-1,n}+\Phi _{m-1,n-m}\right)+\left(\frac{\beta}{2}+\frac{1}{8}\right) \nonumber\\
\fl &&\!\times\psi _1(m+n)+\frac{1}{8}\psi _1(n)+\frac{\beta}{2}  \left(\left(\psi _0(2 n)-\psi _0(m+n)\right){}^2+\psi _1(2 n)-\psi _1(1)\right)\nonumber\\
\fl&& \!+\frac{1}{8} \left(\psi _0(n)-\psi _0(m+n)\right){}^2+\frac{n-m}{2}\left(\psi _0(m+n)-\psi _0(n-m)\right)-m,
\end{eqnarray}
where $\Phi _{a,b}$ is defined in~\eref{eq:sumpsiab} and the coefficient $\beta$ is given by
\begin{equation}
\beta=\frac{(2 m-1) (2 n-1)}{4 m+4 n-2}\label{coef:b0}.
\end{equation}
\end{proposition}

Proposition~\ref{prop2} is proved in~\sref{sec:2.2}. It is important to point out that in deriving the results~\eref{eq:prop1} and~\eref{eq:prop2}, we make use of the lemmas~\ref{lemma1}--\ref{lemma4} in~\cite{HW23} as will also be discussed in~\sref{sec:2.1.2}. The four lemmas are examples of a new simplification framework recently developed in~\cite{HW23} when studying the exact variance of von Neumann entropy. This new framework consists of a set of novel tools useful in simplifying the summations involved, including~\eref{eq:A2S1},~\eref{eq:A2S2}, and~\eref{eq:AA2S} in~appendix A. These summations do not permit further simplifications when using the existing simplification tools for the computation over Hilbert-Schmidt ensemble~\cite{Wei17, Wei20, HWC21} or the Bures-Hall ensemble~\cite{Wei20BHA,Wei20BH,LW21}. For proposition~\ref{prop2}, we also have the following remark.

\begin{remark}
For the same reason as in remark~\ref{remark1}, the result~\eref{eq:prop2} admits closed-form representations for the special cases when the subsystem dimension difference $a=n-m$ is fixed. For example, by fixing $a=0,1,2,3$ in~\eref{eq:prop2}, we recover the recently obtained mean capacity values in~\cite[equations~(27)--(30)]{HW22}.
\end{remark}

Based on the two propositions, the limiting behavior of the average capacity can now be obtained. The results are summarized in corollary~\ref{corollary1} below, and the corresponding proof can be found in~\sref{sec:2.3}.
\begin{corollary}\label{corollary1}
For any subsystem dimensions $m\leq n$ in the asymptotic regime
\begin{equation}\label{regime1}
m\to \infty, ~~~n\to \infty,~~~{with~a~fixed}~n-m,
\end{equation}
the average entanglement capacity of fermionic Gaussian states with a fixed particle number~\eref{eq:prop1} and with an arbitrary particle number~\eref{eq:prop2} approach to the same limit
\begin{equation}\label{eq:corollary1}
\frac{\mathbb{E}\!\left[C\right]}{m}\longrightarrow\frac{\pi ^2}{8}-1. \label{eq:limit1}
\end{equation}
\end{corollary}

In corollary~\ref{corollary1}, we note that for the case of fixed particle number, the particle number $p$ also goes to infinity of the same rate as $m$ and $n$ in the limit~\eref{regime1}. For the case of arbitrary number of particles, the limiting value~\eref{eq:corollary1}, also known as the leading volume-law coefficient, was first obtained in~\cite{BNP21} for equal subsystem dimensions. Here, we have extended it rigorously to a more general regime~\eref{regime1} starting from our explicit result~\eref{eq:prop2}. We also observe the interesting fact that the limiting value~(\ref{eq:limit1}) is the same for the cases~\eref{eq:fgfp} and~\eref{eq:fgap} despite the fundamental difference of the two underlying models.

\begin{figure}[htbp]
\centering
\includegraphics[width=0.85\linewidth]{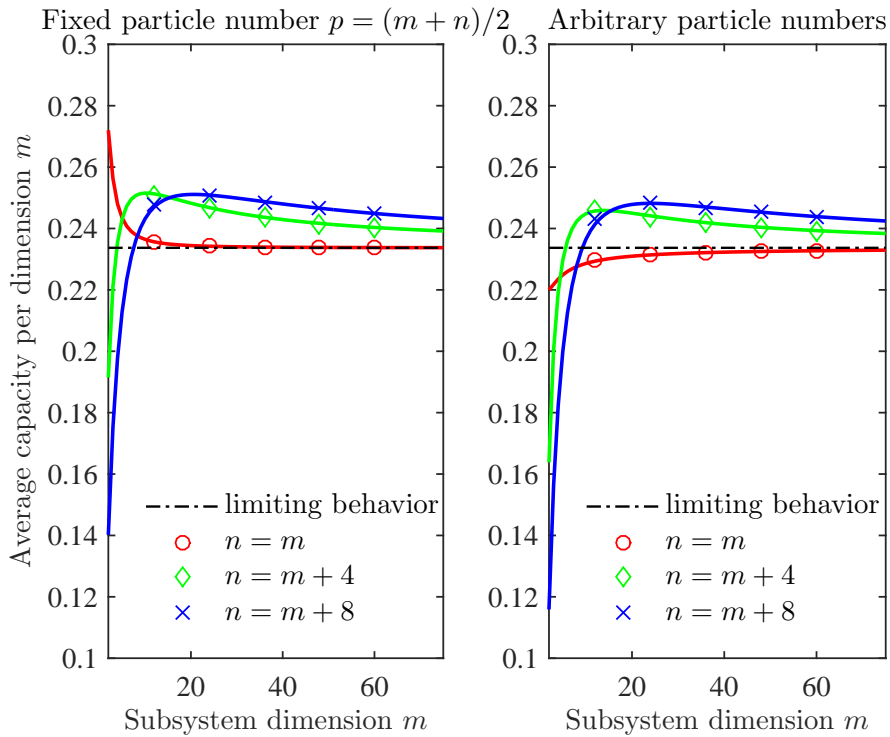}
\caption{Average of entanglement capacity (per dimension) of fermionic Gaussian states with and without particle number constraints: analytical results versus simulations. The solid  lines are drawn by the exact capacity formulas~\eref{eq:prop1} and~\eref{eq:prop2}, while the dash-dot horizontal lines represent the limiting behaviors of average capacity~\eref{eq:limit1}. The corresponding scatters in the symbols of circle, diamond, and asterisk are obtained from numerical simulations.}
\label{fig:p1}
\end{figure}

To illustrate the obtained results, we plot in~\fref{fig:p1} the exact formulas~\eref{eq:prop1} and~\eref{eq:prop2} per dimension $m$ for fixed subsystem dimension differences $n-m=0,~4,~8$, along with the asymptotic value~\eref{eq:limit1}. The left-hand side figure corresponds to the case of a fixed particle number $p=(m+n)/2$, and the right-hand side corresponds to the case of an arbitrary particle number. It is observed that as the dimension difference $n-m$ increases, the average capacity~\eref{eq:prop1} and~\eref{eq:prop2} approach to the limiting value~\eref{eq:limit1} more slowly. This fact indicates that the finite-size capacity formulas are more useful when the dimension difference $n-m$ is large, cf.~\cite{Wei23}, and otherwise the asymptotic value~\eref{eq:limit1} serves as a reasonably accurate approximation. We also plot the simulated values of mean capacity in~\fref{fig:p1}, which match well with the analytical results.

\section{Computation of average capacity}\label{sec:2}
In this section, we prove the results presented in the previous section. The mean formula of entanglement capacity for fermionic Gaussian states with a fixed particle number in proposition~\ref{prop1} is calculated in~\sref{sec:2.1}. The computation for the case of an arbitrary particle number in proposition~\ref{prop2} is performed in~\sref{sec:2.2}. The limiting value of the average capacity in corollary~\ref{corollary1} is proved in~\sref{sec:2.3}.

\subsection{Average capacity over fermionic Gaussian states with particle number constraint}\label{sec:2.1}
Here, we compute the mean value of entanglement capacity~\eref{eq:capacity} over fermionic Gaussian states with particle number constraint~\eref{eq:fgfp}.
The computation mainly consists of two parts. The first part is to obtain a summation representation of the average capacity as shown in~\sref{sec:2.1.1}. In~\sref{sec:2.1.2}, we then simplify the summations in arriving at the desired result~\eref{eq:prop1} in proposition~\ref{prop1}.

\subsubsection{Correlation functions and integral calculations}\label{sec:2.1.1}
Recall the definition~(\ref{eq:capacity}) of entanglement capacity
\begin{equation}
C=\sum_{i=1}^{m}u(x_{i})
\end{equation}
with
\begin{equation}
u(x)=\frac{1-x^{2}}{4}\ln^2\frac{1+x}{1-x},
\end{equation}
computing its average requires the probability density function of one arbitrary eigenvalue of the fermionic Gaussian ensemble. Denoting $g_{l}(x_{1},\dots,x_{l})$ as the joint density of $l$ arbitrary eigenvalues, the average capacity is written as
\begin{equation}\label{eq:EC1}
\mathbb{E}\!\left[C\right]=m\int_{-1}^{1}u(x)g_1(x)\dd x.
\end{equation}
When $\gamma=1$, the ensemble~\eref{eq:fg-ensemble} is the well-known Jacobi unitary ensemble. In this case, the joint density $g_{l}(x_{1},\dots,x_{l})$  can be written in terms of an $l\times l$ determinant as~\cite{Mehta, Forrester}
\begin{equation}\label{eq:gl}
g_{l}(x_{1},\dots,x_{l})=\frac{(m-l)!}{m!}\det\left(K\left(x_{i},x_{j}\right)\right)_{i,j=1}^{l}.
\end{equation}
The determinant in~(\ref{eq:gl}) is known as the $l$-point correlation function~\cite{Forrester}, where
\begin{equation}\label{eq:corrk}
K\left(x,y\right)=\sqrt{w(x)w(y)}\sum_{k=0}^{m-1}\frac{J_k^{(a,b)}(x) J_k^{(a,b)}(y)}{h_k}
\end{equation}
is the correlation kernel with the weight function
\begin{equation}
w(x)=\left(\frac{1-x}{2}\right)^{a}  \left(\frac{1+x}{2}\right)^{b}.
\end{equation}
In~\eref{eq:corrk}, the polynomial $J_k^{(a,b)}(x)$ is the Jacobi polynomial supported in~$x\in[-1,1]$, and
\begin{equation}
 h_{k}=\frac{2\Gamma(k+a+1)\Gamma(k+b+1)}{(2k+a+b+1)\Gamma(k+1)\Gamma(k+a+b+1)}
\end{equation}
is the normalization constant, which is obtained by the orthogonality relation of Jacobi polynomials~\cite{Forrester}
\begin{eqnarray}
&\int_{-1}^{1}\left(\frac{1-x}{2}\right)^{a}\left(\frac{1+x}{2}\right)^{b}J^{(a,b)}_{k}(x)J^{(a,b)}_{l}(x)\dd x\nonumber\\
&=\frac{2\Gamma(k+a+1)\Gamma(k+b+1)}{(2k+a+b+1)\Gamma(k+1)\Gamma(k+a+b+1)}\delta_{kl}, \quad \Re(a,b)>-1.\label{eq:orthogonality}
\end{eqnarray}

By rewriting the function $u(x)$ in~\eref{eq:ux1} as
\begin{eqnarray}\label{eq:ux}
u(x)&=&\frac{1+x}{2}\ln^2\frac{1+x}{2}+\frac{1-x}{2}\ln^2\frac{1-x}{2}\nonumber\\
&&-\left(\frac{1+x}{2}\ln\frac{1+x}{2}+\frac{1-x}{2}\ln\frac{1-x}{2}\right)^2,
\end{eqnarray}
the average capacity~\eref{eq:EC1} boils down to computing two integrals involving the one-point correlation function, cf.~\cite{HW22}, as
\begin{equation}\label{eq:fpac2}
\mathbb{E}\!\left[C\right]=\mathrm{I_{\mathcal{C}}}-\mathrm{I_{\mathcal{A}}},
\end{equation}
where
\begin{eqnarray}
\mathrm{I_{\mathcal{C}}}&=\int_{-1}^{1} \left(\frac{1+x}{2}\ln^{2}\frac{1+x}{2}+\frac{1-x}{2}\ln^{2}\frac{1-x}{2}\right)K(x,x)\dd x \label{eq:IC1}\\
\mathrm{I_{\mathcal{A}}}&=\int_{-1}^{1} \left(\frac{1+x}{2}\ln\frac{1+x}{2}+\frac{1-x}{2}\ln\frac{1-x}{2}\right)^2K(x,x)\dd x\label{eq:IA1}.
\end{eqnarray}
By the definition of the correlation kernel~\eref{eq:corrk}, the integral $\mathrm{I_{\mathcal{C}}}$ in~\eref{eq:IC1} is further written as
\begin{eqnarray}
\mathrm{I_{\mathcal{C}}}=&\sum_{k=0}^{m-1}\frac{1}{h_{k}}\int_{-1}^{1}\left(\frac{1+x}{2}\ln^{2}\frac{1+x}{2}+\frac{1-x}{2}\ln^{2}\frac{1-x}{2}\right)\nonumber\\
 &\times\left(\frac{1-x}{2}\right)^{a}\left(\frac{1+x}{2}\right)^{b}J_{k}^{(a,b)}(x)^2\dd x\label{eq:fpA1ab}.
\end{eqnarray}
Similarly, the integral $\mathrm{I_{\mathcal{A}}}$ in~\eref{eq:IA1} now consists of two parts
\begin{equation}\label{eq:IAex}
\mathrm{I_{\mathcal{A}}}=\mathcal{A}_{1}+\mathcal{A}_{2},
\end{equation}
where
\begin{eqnarray}
\!\!\!\!\!\!\!\! \mathcal{A}_1 &=&\sum_{k=0}^{m-1}\frac{1}{h_{k}}\int_{-1}^{1}\left(\left(\frac{1+x}{2}\right)^{2}\ln^{2}\frac{1+x}{2}+\left(\frac{1-x}{2}\right)^{2}\ln^{2}\frac{1-x}{2}\right)\nonumber\\
&&\times\left(\frac{1-x}{2}\right)^{a}\left(\frac{1+x}{2}\right)^{b}J_{k}^{(a,b)}(x)^2\dd x\\
\!\!\!\!\!\!\!\! \mathcal{A}_2&=&\sum_{k=0}^{m-1}\frac{2}{h_{k}}\int_{-1}^{1}\left(\frac{1-x}{2}\right)^{a+1}\left(\frac{1+x}{2}\right)^{b+1}\ln\frac{1-x}{2}\ln\frac{1+x}{2}J_{k}^{(a,b)}(x)^2\dd x. \label{eq:fpA2}
\end{eqnarray}
Here, we recall that $a=n-p\geq0$ and $b=p-m\geq0$ in~\eref{eq:fgfp}.
Due to the parity property of Jacobi polynomials~\cite{Szego}
\begin{equation}\label{eq:parity}
J^{(a,b)}_{k}(-x)=(-1)^{k}J^{(b,a)}_{k}(x),
\end{equation}
the integrals $\mathrm{I_{\mathcal{C}}}$ and $\mathcal{A}_1$ admit the following symmetric structures
\begin{eqnarray}
\mathrm{I_{\mathcal{C}}}&=&\mathrm{I_{\mathcal{C}}}^{(a,b)}+\mathrm{I_{\mathcal{C}}}^{(b,a)}\label{sym_ic}\\
\mathcal{A}_1&=&\mathcal{A}_1^{(a,b)}+\mathcal{A}_1^{(b,a)}\label{sym_ia1},
\end{eqnarray}
where
\begin{eqnarray}
\mathrm{I_{\mathcal{C}}}^{(a,b)}&=&\sum_{k=0}^{m-1}\frac{1}{h_{k}}\int_{-1}^{1}\left(\frac{1-x}{2}\right)^{a}\left(\frac{1+x}{2}\right)^{b+1}\ln^{2}\frac{1+x}{2}J_{k}^{(a,b)}(x)^2\dd x\label{eq:fpIcab}\\
\mathcal{A}_1^{(a,b)}&=&\sum_{k=0}^{m-1}\frac{1}{h_{k}}\int_{-1}^{1}\left(\frac{1-x}{2}\right)^{a}\left(\frac{1+x}{2}\right)^{b+2}\ln^{2}\frac{1+x}{2}J_{k}^{(a,b)}(x)^2\dd x.\label{eq:fpA1ab}
\end{eqnarray}
The summations in~(\ref{eq:fpA2}),~\eref{eq:fpIcab}, and (\ref{eq:fpA1ab}) can be evaluated by using the confluent form of Christoffel-Darboux formula~\cite{Forrester}
\begin{equation}\label{eq:cd}
\sum_{k=0}^{m-1}\frac{J_k^{(a,b)}(x)^2}{h_k}=\alpha_1 J_{m-1}^{(a+1,b+1)}(x)
J_{m-1}^{(a,b)}(x)- \alpha_2 J_{m-2}^{(a+1,b+1)}(x) J_m^{(a,b)}(x),
\end{equation}
where
\begin{eqnarray}
\alpha_1&=\frac{m (a+b+m)(a+b+m+1)}{h_{m-1} (a+b+2 m-1)_2}\\
\alpha_2&=\frac{m (a+b+m)^2}{h_{m-1} (a+b+2 m-1)_2}.
\end{eqnarray}
Consequently, we have
\begin{eqnarray}
\fl \mathrm{I_{\mathcal{C}}}^{(a,b)}&=&\alpha_1\int_{-1}^1 \left(\frac{1-x}{2}\right)^a \left(\frac{1+x}{2}\right)^{b+1} \ln ^2\frac{1+x}{2}J_{m-1}^{(a+1,b+1)}(x) J_{m-1}^{(a,b)}(x) \dd x\nonumber\\
\fl &&\!-\alpha_2\int_{-1}^1 \left(\frac{1-x}{2}\right)^a \left(\frac{1+x}{2}\right)^{b+1} \ln ^2\frac{1+x}{2}J_{m-2}^{(a+1,b+1)}(x) J_{m}^{(a,b)}(x) \dd x
\label{eq:Iccd}\\
\fl \mathcal{A}_1^{(a,b)}&=&\alpha_1\int_{-1}^1 \left(\frac{1-x}{2}\right)^a \left(\frac{1+x}{2}\right)^{b+2} \ln ^2\frac{1+x}{2}J_{m-1}^{(a+1,b+1)}(x) J_{m-1}^{(a,b)}(x) \dd x\nonumber\\
\fl &&\!-\alpha_2\int_{-1}^1 \left(\frac{1-x}{2}\right)^a \left(\frac{1+x}{2}\right)^{b+2} \ln ^2\frac{1+x}{2}J_{m-2}^{(a+1,b+1)}(x) J_{m}^{(a,b)}(x) \dd x
\label{eq:A1cd}
\end{eqnarray}
and
\begin{equation}\label{eq:A2}
\mathcal{A}_2=2\alpha_1\mathcal{A}_2(m-1,m-1)-2\alpha_2\mathcal{A}_2(m-2,m),
\end{equation}
where
\begin{eqnarray}
\fl\mathcal{A}_2(m-1,m-1)&=&\int_{-1}^1 \left(\frac{1-x}{2}\right)^{a+1} \left(\frac{1+x}{2}\right)^{b+1}\nonumber\\
\fl &&\times \ln \frac{1-x}{2}\ln \frac{1+x}{2} J_{m-1}^{(a+1,b+1)}(x) J_{m-1}^{(a,b)}(x) \dd x \label{eq:a2m-1}\\
\fl~~\!~~~~\mathcal{A}_2(m-2,m)&=&\int_{-1}^1 \left(\frac{1-x}{2}\right)^{a+1} \left(\frac{1+x}{2}\right)^{b+1} \nonumber\\
\fl &&\times\ln \frac{1-x}{2}\ln \frac{1+x}{2} J_{m-2}^{(a+1,b+1)}(x) J_{m}^{(a,b)}(x)\label{eq:a2m-2} \dd x.
\end{eqnarray}
Computing the above integrals $\mathrm{I_{\mathcal{C}}}^{(a,b)}$ and $\mathcal{A}_1^{(a,b)}$ in~\eref{eq:Iccd}--\eref{eq:A1cd} requires the integral identity
\begin{eqnarray}
&\int_{-1}^{1}\left(\frac{1-x}{2}\right)^{a_1}\left(\frac{1+x}{2}\right)^{c}J^{(a_1,b_1)}_{k_1}(x)J^{(a_2,b_2)}_{k_2}(x)\dd x \nonumber \\
&=\frac{2 \left(k_1+1\right)_{a_1}}{\left(b_2+k_2+1\right)_{a_2}}\sum _{i=0}^{k_2} \frac{(-1)^{i+k_2} (i+1)_c \left(i+b_2+1\right)_{a_2+k_2} }{\Gamma \left(k_2-i+1\right) \Gamma \left(a_1+c+i+k_1+2\right)}\nonumber\\
&~~~\!~\times\left(c+i-b_1-k_1+1\right)_{k_1}, \quad \Re(a_1,a_2,b_1,b_2,c)>-1.\label{eq:SIac2}
\end{eqnarray}
To show this identity, we first note that the Jacobi polynomial $J_k^{(a,b)}(x)$ supported in $x\in[-1,1]$ admits different representations~\cite{Szego, Forrester}
\begin{eqnarray}\label{eq:J1}
\!\!\!\!\!\!\!\!\!\!\!\!J^{(a,b)}_{k}(x)&=&\frac{(-1)^{k}(b+1)_{k}}{k!}\sum_{i=0}^{k}\frac{(-k)_{i}(k+a+b+1)_{i}}{(b+1)_{i}\Gamma(i+1)}\left(\frac{1+x}{2}\right)^{i}\\
\!\!\!\!\!\!\!\!\!\!\!\!&=&\sum _{i=0}^k \frac{(-1)^i \Gamma (a+k+1) (k+b-i+1)_i}{\Gamma (i+1) \Gamma (a+i+1) \Gamma (k-i+1)}\left(\frac{1-x}{2}\right)^i \left(\frac{1+x}{2}\right)^{k-i}. \label{eq:J2}
\end{eqnarray}
The identity~(\ref{eq:SIac2}) is then obtained by using the definition~(\ref{eq:J1}) for the polynomial $J^{(a_2,b_2)}_{k_2}$ before applying the well-known integral identity~\cite{Szego, Forrester}
\begin{eqnarray}
&\int_{-1}^{1}\left(\frac{1-x}{2}\right)^{a}\left(\frac{1+x}{2}\right)^{c}J^{(a,b)}_{k}(x)\dd x \nonumber \\
&=\frac{2 \Gamma (c+1) (k+1)_a (c-b-k+1)_k}{\Gamma (a+c+k+2)}, \quad \Re(a,b,c)>-1.\label{eq:SIac}
\end{eqnarray}
In~\eref{eq:SIac2}, by specializing
\begin{equation}\label{eq:sp1}
a_1=a,~~~a_2=a+1,~~~b_1=b,~~~b_2=b+1,~~~k_1=k_2=m-1
\end{equation}
so that
\begin{equation}
 J^{(a_1,b_1)}_{k_1}(x)\to J_{m-1}^{(a,b)}(x),\qquad J^{(a_2,b_2)}_{k_2}(x)\to J_{m-1}^{(a+1,b+1)}(x),
\end{equation}
the first integral in~\eref{eq:Iccd} can now be computed by taking twice derivatives with respect to the parameter $c$ of the specialized identity~\eref{eq:SIac2} before setting $c=b+1$. Other integrals in~\eref{eq:Iccd}--\eref{eq:A1cd} are calculated in the same manner.

To compute the integral~$\mathcal{A}_2$ in~\eref{eq:A2}, one will need another integral identity
\begin{eqnarray}
&\int_{-1}^{1}\left(\frac{1-x}{2}\right)^{d}\left(\frac{1+x}{2}\right)^{c}J^{(a_1,b_1)}_{k_1}J^{(a_2,b_2)}_{k_2}(x)\dd x \nonumber\\
&=\frac{2 \Gamma \left(a_2+k_2+1\right) \Gamma \left(b_2+k_2+1\right)}{\Gamma \left(c+d+k_1+k_2+2\right)}\sum _{i=0}^{k_2} \frac{(-1)^i \Gamma \left(d-a_1+i+1\right)}{\Gamma (i+1) \Gamma \left(a_2+i+1\right)}\nonumber\\
&~~~\!~\times\frac{\Gamma \left(c-b_1-i+k_2+1\right)}{\Gamma \left(k_2-i+1\right) \Gamma \left(b_2-i+k_2+1\right)}\sum _{j=0}^{k_1} \frac{(-1)^j \left(k_1-j+1\right)_{d+i}}{\Gamma (j+1)}\nonumber\\
&~~~\!~\times\frac{\left(c-i+j-b_1-k_1+k_2+1\right)_{b_1+k_1}}{\Gamma \left(d-a_1+i-j+1\right)}, \quad \Re(a_1,a_2,b_1,b_2,c,d)>-1,\label{eq:SIdc2}
\end{eqnarray}
which is obtained by using the definition~(\ref{eq:J2}) for the polynomial $J^{(a_2,b_2)}_{k_2}$  before applying the identity~\cite[equation (62)]{HW22}
\begin{eqnarray}
&\int_{-1}^{1}\left(\frac{1-x}{2}\right)^{d}\left(\frac{1+x}{2}\right)^{c}J^{(a,b)}_{k}(x)\dd x  \nonumber \\
&=\frac{2 \Gamma (c-b+1) \Gamma (d-a+1)}{\Gamma (c+d+k+2)}\sum _{i=0}^{k} \frac{(-1)^i \Gamma (c+i+1) \Gamma (d-i+k+1)}{\Gamma (i+1) \Gamma (k-i+1)}\nonumber\\
&~~~\!~\times\frac{1}{ \Gamma (d-a-i+1) \Gamma (c-b+i-k+1)}, \quad \Re(a,b,c,d)>-1\label{eq:SIcd}.
\end{eqnarray}
The two integrals~\eref{eq:a2m-1} and \eref{eq:a2m-2} in $\mathcal{A}_2 $ are calculated by taking derivatives of $c$ and $d$ of identity~(\ref{eq:SIdc2}) with the specialization~(\ref{eq:sp1}) and the specialization
\begin{equation}\label{eq:sp2}
a_1=a,~~~b_1=b,~~~a_2=a+1,~~~b_2=b+1,~~~k_1=m,~~~k_2=m-2,
\end{equation}
respectively, before setting $c=b+1$, $d=a+1$.

In writing down the summation forms of $\mathrm{I_{\mathcal{C}}}^{(a,b)}$, $\mathcal{A}_1^{(a,b)}$, and $\mathcal{A}_2^{(a,b)}$, one will also have to resolve the indeterminacy by using the following asymptotic expansions of gamma and polygamma functions of negative arguments~\cite{AS72} when $\epsilon \rightarrow 0$,
\label{eq:pgna}\begin{eqnarray}
\Gamma(-l+\epsilon)&=\frac{(-1)^{l}}{l!\epsilon}\left(1+\psi_{0}(l+1)\epsilon+o\left(\epsilon^2\right)\right)\label{eq:pgna1}\\
\psi_{0}(-l+\epsilon)&=-\frac{1}{\epsilon}+\psi_{0}(l+1)+\left(2\psi_{1}(1)-\psi_{1}(l+1)\right)\epsilon+o\left(\epsilon^2\right)\label{eq:pgna2}\\
\psi_{1}(-l+\epsilon)&=\frac{1}{\epsilon^{2}}-\psi_{1}(l+1)+\psi_{1}(1)+\zeta(2)+o\left(\epsilon\right).\label{eq:pgna3}
\end{eqnarray}

The resulting summation forms of $\mathrm{I_{\mathcal{C}}}^{(a,b)}$, $\mathcal{A}_1^{(a,b)}$, and $\mathcal{A}_2$ are summarized in~\eref{eq:IcabS}--\eref{eq:A2S2} in~appendix A.1.

\subsubsection{Simplification of summations} \label{sec:2.1.2}
The remaining task in computing the average capacity
\begin{equation}
\mathbb{E}\!\left[C\right]=\mathrm{I_{\mathcal{C}}}-\mathrm{I_{\mathcal{A}}},
\end{equation}
is to simplify the summations in~(\ref{eq:IcabS})--(\ref{eq:A2S2}). In the subsequent calculation, we first simplify the summation~(\ref{eq:IcabS}) in obtaining $\mathrm{I_{\mathcal{C}}}$, whereas $\mathrm{I_{\mathcal{A}}}$ is obtained by simplifying the summations \eref{eq:A1abS}--\eref{eq:A2S2}.

We first simplify the summations in~\eref{eq:IcabS}. Note that the first two sums in~\eref{eq:IcabS} are single sums consisting of polygamma and rational functions, and the last sum can be directly reduced to a closed-form expression. The two single summations are simplified, by using the identities~\eref{eq:B1}--\eref{eq:B8} while keeping in mind the symmetric structure~\eref{sym_ic}
\begin{equation}\label{result:ic}
\mathrm{I_{\mathcal{C}}}=\mathrm{I_{\mathcal{C}}^{(a,b)}}+\mathrm{I_{\mathcal{C}}^{(b,a)}},
\end{equation}
as
\begin{eqnarray}\label{result:icab}
\fl \mathrm{I_{\mathcal{C}}^{(a,b)}}&=&a_0 \sum _{k=1}^m \frac{\psi _0(a+b+k+m)}{b+k}-a_1 \sum _{k=1}^m \frac{\psi _0(a+b+k+m)}{k}+a_1 \sum _{k=1}^m \frac{\psi _0(b+k)}{k}+a_2 \nonumber\\
\fl&&\times\bigg(\psi_0^{2}(a+b+2 m)-\psi _0(a+b+m) \psi _0(a+b+2 m)-\psi _0(a+b+2 m)\nonumber\\
\fl&&\times\psi _0(b+m) \bigg)+a_0 \psi _0(b) \psi _0(a+b+m)+\frac{a_1}{2}\bigg(\psi _1(b)-\psi _1(a+b+m)\nonumber\\
\fl&&+\psi _0(a+b+m) \left(\psi _0(a+b+m)+2 \psi _0(m)-2 \psi _0(1)\right)+2 \psi _0(b) (\psi _0(b+m)\nonumber\\
\fl&&-\psi _0(m)+\psi _0(1))-\psi_0^{2}(b)\bigg)+a_3 \psi_0(a+b+2 m)+a_4 \psi _0(a+b+m)\nonumber\\
\fl&&+a_5 \psi _0(b+m)+a_6 \psi _0(b)+a_7,
\end{eqnarray}
where the coefficients $a_i$ are summarized in~\eref{coeff_a0}--\eref{coeff_a7} of~appendix C.1.

We now simplify the summations~\eref{eq:A1abS}--\eref{eq:A2S2} in obtaining $\mathrm{I_{\mathcal{A}}}$. The summation~\eref{eq:A1abS} is simplified into a similar form as the result~\eref{result:icab} by using the identities~\eref{eq:B1}--\eref{eq:B8}. The integral~${\mathcal{A}}_1$ is then obtained by adding the result of~\eref{eq:A1abS} and its symmetric form according to~\eref{sym_ia1}. Continue to simplify the summations~\eref{eq:A2S1} and~\eref{eq:A2S2} will require the following four lemmas.\\

\begin{lemma}\label{lemma1}
For any complex numbers $a,b,c\notin \mathbb{Z}^{-}$,
we have
\begin{eqnarray}
\fl& \sum _{i=1}^m \frac{1}{\Gamma (i) \Gamma (a+i) \Gamma (m+1-i) \Gamma (m+b+1-i)(c+i) }\nonumber \\
 \fl&=\frac{1}{\Gamma (b+m) \Gamma (c+m+1) \Gamma (a+b+m)} \sum _{i=1}^m \frac{\Gamma (c-i+m+1) \Gamma (a+b-i+2 m)}{\Gamma (m-i+1) \Gamma (a-i+m+1)}.\label{eq:lemma1}\\ \fl\nonumber
\end{eqnarray}
\end{lemma}

\begin{lemma}\label{lemma2}
For any complex numbers $a,b\notin \mathbb{Z^{-}}$, and any $c\in \mathbb{Z^{+}}$, we have
\begin{eqnarray}
\fl&\sum _{i=1}^m \frac{1}{\Gamma (c+i) \Gamma (a+i) \Gamma (m+1-i) \Gamma (m+b+1-i) }\nonumber\\
\fl&=\frac{1}{\Gamma (m+b) \Gamma (m+a+b) \Gamma (c) \Gamma (m+c)}\sum _{i=1}^m \frac{\Gamma (m+a+b+i-1) \Gamma (m+c-i)}{\Gamma (a+i) \Gamma (m-i+1)}\label{eq:lemma2}.\\ \fl\nonumber
\end{eqnarray}
\end{lemma}

\begin{lemma}\label{lemma3}
For any complex numbers $a,b\notin \mathbb{Z^{-}}$, and any $c\in \mathbb{Z^{+}}$, we have
\begin{eqnarray}
\fl &\sum _{i=1}^m \frac{1}{\Gamma (c+i) \Gamma (a+i) \Gamma (m-i+1) \Gamma (b-i+m+1) i }\nonumber \\
\fl &=\frac{1}{\Gamma (a) \Gamma (a+m) \Gamma (1+b+m) \Gamma (b+c+m)}\sum _{i=1}^m \frac{\Gamma (a-i+m) \Gamma (b+c+i+m)}{ \Gamma (c+i) \Gamma (m-i+1)i}\nonumber\\
\fl &~~~\!~+\frac{\psi_0 (a)-\psi_0 (a+m)}{\Gamma (a) \Gamma (c) \Gamma (m+1) \Gamma (b+m+1)}.\label{eq:lemma3}\\ \fl\nonumber
\end{eqnarray}
\end{lemma}

\begin{lemma}\label{lemma4}
For any complex numbers $a,b\notin \mathbb{Z^{-}}$, and any $c, d\in \mathbb{Z^{+}}$, we have
\begin{eqnarray}
\fl &\sum _{i=1}^m \frac{1}{\Gamma (c+i) \Gamma (a+i) \Gamma (d+m-i+1) \Gamma (b+m-i+1)}\nonumber\\
\fl &=\frac{1}{\Gamma (d) \Gamma (a+m) \Gamma (a+b+m) \Gamma (c+d+m)}\sum _{i=1}^m \frac{\Gamma (c+d+i-1) \Gamma (a+b-i+2 m)}{\Gamma (c+i) \Gamma (b-i+m+1)}\nonumber\\
\fl &~~~+\frac{1}{\Gamma (c) \Gamma (b+m) \Gamma (a+b+m) \Gamma (c+d+m)}\sum _{i=1}^m \frac{\Gamma (c+d+i-1) \Gamma (a+b-i+2 m)}{\Gamma (d+i) \Gamma (a-i+m+1)}.\nonumber\\
\fl
\label{eq:lemma4}
\end{eqnarray}
\end{lemma}
Proof to the above four lemmas is based on a new simplification framework proposed in~\cite[section 2.2.2]{HW23}. Equipped with these tools, the summations~\eref{eq:A2S1} and~\eref{eq:A2S2} can now be simplified. In the following, we will first present the simplification of \eref{eq:A2S2}, whereas \eref{eq:A2S1} is simplified in the same manner.

Note that~\eref{eq:A2S2} consists of one single summation and two double summations. To proceed with the single summation
\begin{eqnarray}\label{eq:A2SS0}
\fl&\sum _{i=1}^{m-1} \frac{1}{\Gamma (i) \Gamma (a+i+1) \Gamma (m-i) \Gamma (b-i+m+1)}\bigg((\psi _0(a+b+2 m+2)-\psi _0(a+m+1)\nonumber\\
\fl&-\psi _0(i+1)+\psi _0(1)) (\psi _0(m-i+1)-\psi _0(a+b+2 m+2)+\psi _0(b+m+1)-\psi _0(1))\nonumber\\
\fl&+\psi _1(a+b+2 m+2)\bigg),
\end{eqnarray}
we first rewrite it as
\begin{eqnarray}\label{eq:A2SS1}
\left(s_0-s_1 s_2\right) \sum _{i=1}^{m-1} \frac{1}{\Gamma (i) \Gamma (a+i+1) \Gamma (m-i) \Gamma (b-i+m+1)}\nonumber\\
+\left(s_1-\frac{1}{m}\right) \sum _{i=1}^{m-1} \frac{1}{\Gamma (i) \Gamma (a+i+1) \Gamma (m-i+1) \Gamma (b-i+m+1)}\nonumber\\
+\left(s_2-\frac{1}{m}\right) \sum _{i=1}^{m-1} \frac{1}{\Gamma (i+1) \Gamma (a+i+1) \Gamma (m-i) \Gamma (b-i+m+1)}
\nonumber\\
+s_1 \sum _{i=1}^{m-1} \frac{\psi _0(i)}{\Gamma (i) \Gamma (b+i+1) \Gamma (m-i) \Gamma (a-i+m+1)}
\nonumber\\
-\sum _{i=1}^{m-1} \frac{\psi _0(i)}{\Gamma (i) \Gamma (b+i+1) \Gamma (m-i+1) \Gamma (a-i+m+1)}
\nonumber\\
-\sum _{i=1}^{m-1} \frac{\psi _0(i)}{\Gamma (i) \Gamma (a+i+1) \Gamma (m-i+1) \Gamma (b-i+m+1)}
\nonumber\\
+s_2 \sum _{i=1}^{m-1} \frac{\psi _0(i)}{\Gamma (i) \Gamma (a+i+1) \Gamma (m-i) \Gamma (b-i+m+1)}\nonumber\\
-\sum _{i=1}^{m-1} \frac{\psi _0(i) \psi _0(m-i)}{\Gamma (i) \Gamma (a+i+1) \Gamma (m-i) \Gamma (b-i+m+1)},
\end{eqnarray}
where
\begin{eqnarray}
s_0&=&\psi _1(a+b+2 m+2)\\
s_1&=&\psi _0(a+b+2 m+2)-\psi _0(a+m+1)+\psi _0(1)\\
s_2&=&\psi _0(a+b+2 m+2)-\psi _0(b+m+1)+\psi _0(1).
\end{eqnarray}
The summations in~\eref{eq:A2SS1} are then simplified into single sums of the forms
\begin{eqnarray}
\sum _{j=1}^{m} \frac{\Gamma (a+b-j+2 m-1)}{ \Gamma (a-j+m)j}\label{basis1}\\
\sum _{j=1}^{m} \frac{\Gamma (a+b-j+2 m-1)}{\Gamma (a-j+m)j^2 }\label{basis2}
\end{eqnarray}
by using lemma~\ref{lemma2},~lemma \ref{lemma4}, and the closed-form identity~\cite{Luke}
\begin{eqnarray}
&\sum _{i=1}^m \frac{1}{\Gamma (i) \Gamma (a+i) \Gamma (m-i+1) \Gamma (m+b+1-i)}\nonumber\\
&=\frac{\Gamma (a+b+2m-1)}{\Gamma (m) \Gamma (a+m) \Gamma (b+m) \Gamma (a+b+m) }. \label{eq:ofgi}
\end{eqnarray}
More specifically, the first three summations in~\eref{eq:A2SS1} are simplified into closed-form expressions by using the identity~\eref{eq:ofgi}, and the next four summations are simplified by taking derivative of $c$ of the identity~\eref{eq:lemma2} in lemma~\ref{lemma2} before setting $c=0$. The last summation in~\eref{eq:A2SS1} is simplified by taking derivatives of $c$ and $d$ of the identity~\eref{eq:lemma4} in lemma~\ref{lemma4} before setting $c=d=0$.

We now move on to the double summations in~\eref{eq:A2S2}, which are
\begin{eqnarray}\label{eq:A2S2a}
\!\!\!\!\!\!\!\!\!\!\!\!\!\!\!\!&\sum _{i=1}^{m-1} \frac{i (m-i)}{\Gamma (b+i+1) \Gamma (a-i+m+1)}\sum _{j=1}^{m-i} \frac{\Gamma (a+j+m+1) \Gamma (b-j+m+1)}{j \Gamma (i+j+1) \Gamma (m-i-j+1)}\nonumber\\
\!\!\!\!\!\!\!\!\!\!\!\!\!\!\!\! &\times\left(\psi _0(a+j+m+1)-\psi _0(a+b+2 m+2)+\psi _0(m-i+1)-\psi _0(j+1)\right)
\end{eqnarray}
and
\begin{eqnarray}\label{eq:A2S2b}
\!\!\!\!\!\!\!\!\!\!\!\!\!\!\!\!&\sum _{i=1}^{m-1} \frac{i (m-i)}{\Gamma (a+i+1) \Gamma (b-i+m+1)}\sum _{j=1}^{m-i} \frac{\Gamma (a-j+m+1) \Gamma (b+j+m+1)}{j \Gamma (i+j+1) \Gamma (m-i-j+1)}\nonumber\\
\!\!\!\!\!\!\!\!\!\!\!\!\!\!\!\! &\times(\psi _0(b+j+m+1)-\psi _0(a+b+2 m+2)+\psi _0(m-i+1)-\psi _0(j+1)).
\end{eqnarray}
The two summations~\eref{eq:A2S2a} and~\eref{eq:A2S2b} admit a similar symmetric structure as~\eref{sym_ic}--\eref{sym_ia1}. Therefore, by simplifying the summation~\eref{eq:A2S2a}, the summation~\eref{eq:A2S2b} can be directly obtained by switching $a$ and $b$. We start with the summation~\eref{eq:A2S2a} by dividing it into two parts
\begin{eqnarray}\label{eq:a2d1}
\!\!\!\!\!\!\!\!\!\!\!\!\!\!\!\!&\sum _{i=1}^{m-1} \frac{i (m-i)}{\Gamma (b+i+1) \Gamma (a-i+m+1)}\sum _{j=1}^{m-i} \frac{\Gamma (a+j+m+1) \Gamma (b-j+m+1)}{j \Gamma (i+j+1) \Gamma (m-i-j+1)}\nonumber\\
\!\!\!\!\!\!\!\!\!\!\!\!\!\!\!\! &\times\left(-\psi _0(a+b+2 m+2)-\psi _0(j+1)\right)
\end{eqnarray}
and
 \begin{eqnarray}\label{eq:a2d2}
\!\!\!\!\!\!\!\!\!\!\!\!\!\!\!\!&\sum _{i=1}^{m-1} \frac{i (m-i)}{\Gamma (b+i+1) \Gamma (a-i+m+1)}\sum _{j=1}^{m-i} \frac{\Gamma (a+j+m+1) \Gamma (b-j+m+1)}{j \Gamma (i+j+1) \Gamma (m-i-j+1)}\nonumber\\
\!\!\!\!\!\!\!\!\!\!\!\!\!\!\!\! &\times\left(\psi _0(m-i+1)+\psi _0(a+j+m+1)\right).
\end{eqnarray}
In~\eref{eq:a2d1}, after changing the summation order as
\begin{eqnarray}
\!\!\!\!\!\!\!\!&\sum _{j=1}^{m-1} \frac{\Gamma (a+j+m+1) \Gamma (b-j+m+1)}{j} \left(-\psi _0(a+b+2 m+2)-\psi _0(j+1)\right)\nonumber\\
\!\!\!\!\!\!\!\!&\times\sum _{i=1}^{m-j} \frac{i (m-i)}{\Gamma (b+i+1) \Gamma (i+j+1) \Gamma (a-i+m+1) \Gamma (m-i-j+1)},
\end{eqnarray}
 we evaluate the sum over $i$ by using~lemma~\ref{lemma2}. The double becomes
\begin{eqnarray}\label{eq:a2d1next}
\fl &\frac{1}{\Gamma (b) \Gamma (a+m)}\Bigg(\!(1-a-m) \sum _{j=1}^{m-1} \frac{(a+j+m) (b-j+m) }{j}(\psi _0(a+b+2 m+2) \nonumber\\
\fl &+\psi _0(j+1))\times\sum _{i=1}^{m-j} \frac{\Gamma (b+i-1) \Gamma (a-i+2 m)}{\Gamma (i) \Gamma (m-i+2)}+\frac{a (a+m)}{a+b+m}\sum _{j=1}^{m-1} \frac{b-j+m }{j}(\psi _0(j+1)\nonumber\\
\fl &+\psi _0(a+b+2 m+2))\sum _{i=1}^{m-j} \frac{\Gamma (b+i-1) \Gamma (a-i+2 m+1)}{\Gamma (i) \Gamma (m-i+2)}+\frac{(a+m-1) (b+m)}{a+b+m}\nonumber\\
\fl &\times\sum _{j=1}^{m-1} \frac{(a+j+m)}{j} (\psi _0(a+b+2 m+2)+\psi _0(j+1))\sum _{i=1}^{m-j} \frac{\Gamma (b+i) \Gamma (a-i+2 m)}{\Gamma (i) \Gamma (m-i+2)}\Bigg),
\end{eqnarray}
where the sums over $j$ can be further simplified into closed-form expressions by using the identity~\eref{eq:B3}. As a result, the remaining summations only involve single sums as in~\eref{eq:A2SS1} that are simplified similarly.

The sum~\eref{eq:a2d2} is simplified by first using lemma~\ref{lemma3} along with its derivative with respect to $b$ to evaluate the inner sum over $j$. As a result, the remaining sums are reduced to single sums after computing the sum over $i$ except for the sum
\begin{eqnarray}\label{eq:a2d2next}
\sum _{j=1}^m \frac{1}{\Gamma (j-1) \Gamma (a+j) \Gamma (m-j+1) \Gamma (b-j+m+2)}\nonumber\\
\times\sum _{i=1}^{m-j+1} \left(\frac{\psi_0(a+i+j)}{i}+\frac{\psi_0(i+j)}{i}\right).
\end{eqnarray}
To proceed with~(\ref{eq:a2d2next}), we first use the identity~\eref{eq:B9} to compute the inner sum
\begin{equation}
\sum _{i=1}^{m-j+1} \frac{\psi _0(a+i+j)}{i}
\end{equation}
into
\begin{eqnarray}\label{eq:a2d2next1}
\!\!\!\!\!\!\!\!\!\!\!\!\!\!\!\!\!\!&&\sum _{i=1}^{m-j+1} \frac{\psi _0(i+j)}{i}-\sum _{l=1}^a \frac{\psi _0(l+m+1)}{j+l-1}+\frac{1}{2} \bigg(\!\left(\psi _0(a+j)-\psi _0(j)\right) \nonumber\\
\!\!\!\!\!\!\!\!\!\!\!\!\!\!\!\!\!\!&&\times\left(\psi _0(a+j)+2 \psi _0(m-j+2)+\psi _0(j)-2 \psi _0(1)\right)-\psi _1(a+j)+\psi _1(j)\bigg).
\end{eqnarray}
Inserting the result~\eref{eq:a2d2next1} into~\eref{eq:a2d2next}, the double sum in~\eref{eq:a2d2next} now boils down to simplifying the three sums
\begin{eqnarray}\label{eq:a2d2next2}
\fl&\frac{1}{2} \sum _{j=1}^m \frac{1}{\Gamma (j-1) \Gamma (a+j) \Gamma (m-j+1) \Gamma (b-j+m+2)}\bigg(\psi _1(j)+\left(\psi _0(a+j)-\psi _0(j)\right)\nonumber\\
\fl & \left(\psi _0(a+j)+2 \psi _0(m-j+2)+\psi _0(j)-2 \psi _0(1)\right)-\psi _1(a+j)\bigg),
\end{eqnarray}
\begin{eqnarray}\label{eq:a2d2next3}
\fl\sum _{l=1}^a \psi_0(l+m+1)\sum_{j=1}^m \frac{-1}{\Gamma (j-1) \Gamma (a+j) \Gamma (m-j+1) \Gamma (b-j+m+2)(j+l-1)},
\end{eqnarray}
and
\begin{eqnarray}\label{eq:a2d2next4a}
 \fl \sum _{j=1}^m \frac{2}{\Gamma (j-1) \Gamma (a+j) \Gamma (m-j+1) \Gamma (b-j+m+2)}\sum _{i=1}^{m-j+1} \frac{\psi_0(i+j)}{i}.
\end{eqnarray}
The single sum~\eref{eq:a2d2next2} is simplified in the same manner as~\eref{eq:A2SS1}. For the double summation in~\eref{eq:a2d2next3}, after evaluating the inner sum over $j$ by using lemma~\ref{lemma1}, we arrive at
\begin{eqnarray}\label{eq:a2d2next5}
&-\frac{1}{\Gamma (b+m) \Gamma (a+b+m+1)}\sum _{l=1}^a \frac{\psi _0(l+m+1)}{\Gamma (l+m)}\nonumber\\&\times\sum _{j=1}^{m-1} \frac{\Gamma (m-j+l) \Gamma (a+b-j+2 m)}{\Gamma (m-j) \Gamma (a-j+m+1)}.
\end{eqnarray}
The above sum~\eref{eq:a2d2next5} can now be simplified into single sums by using the identities~\eref{eq:B13}--\eref{eq:B14} to evaluate the sum over $l$, where the remaining single sums are
\begin{eqnarray}
\sum _{j=1}^{m} \frac{\Gamma (a+b-j+2 m-1)}{ \Gamma (a-j+m)j}\psi_0(a+b-j+2 m-1)\label{basis3},
\end{eqnarray}
and
\begin{eqnarray}
\sum _{j=1}^{m} \frac{\psi_0(a+b+j+m)}{j}\label{basis4}.
\end{eqnarray}
So far, the only part that remains to be simplified in~\eref{eq:A2S2a} is the double sum~\eref{eq:a2d2next4a}. We first point out that the sum~\eref{eq:a2d2next4a} has to be treated together with its symmetric part in~\eref{eq:A2S2b}, which is
\begin{eqnarray}\label{eq:a2d2next4b}
\!\!\!\!\!\!\!\! \sum_{j=1}^m \frac{2}{\Gamma (j-1) \Gamma (b+j) \Gamma (m-j+1) \Gamma (a-j+m+2)}\sum _{i=1}^{m-j+1} \frac{\psi_0(i+j)}{i}.
\end{eqnarray}
The two summations~\eref{eq:a2d2next4a} and~\eref{eq:a2d2next4b} may not be further simplified individually. However, we observe cancellations among the two sums by adding them up, where the key ingredient is the identity~\eref{eq:B9}. Specifically, we evaluate the inner summation
\begin{equation}
\sum _{i=1}^{m-j+1} \frac{\psi_0 (i+j)}{i}
\end{equation}
in~\eref{eq:a2d2next4a} by the identity~\eref{eq:B9} with the specialization
\begin{equation}
a\to j,\qquad b\to 0,\qquad m\to m-j+1.
\end{equation}
The sum~\eref{eq:a2d2next4a} becomes
\begin{eqnarray}\label{eq:a2d2next4acancel}
\fl-\sum _{j=1}^m \frac{2}{\Gamma (j-1) \Gamma (a+j) \Gamma (m-j+1) \Gamma (b+m-j+2)}\sum _{i=1}^{j-1} \frac{\psi _0(m-j+i+2)}{i}\nonumber\\
\fl+\sum _{j=1}^m \frac{1}{\Gamma (j-1) \Gamma (a+j) \Gamma (m-j+1) \Gamma (b+m-j+2)}\bigg(\!\left(\psi _0(m-j+2)+\psi _0(j)\right)\nonumber\\
\fl\times\left(\psi _0(m-j+2)+\psi _0(j)-2 \psi _0(1)\right) -\psi _1(m-j+2)-\psi _1(j)+2 \psi _1(1)\bigg).
\end{eqnarray}
Shifting the index $j \to m+2-j$ of the double sum in~\eref{eq:a2d2next4acancel} as
\begin{eqnarray}\label{eq:a2d2next4acancel1}
\fl-\sum _{j=1}^m \frac{2}{\Gamma (j-1) \Gamma (a+j) \Gamma (m-j+1) \Gamma (b+m-j+2)}\sum _{i=1}^{j-1} \frac{\psi _0(m-j+i+2)}{i}\nonumber\\
\fl =-\sum _{j=2}^{m+1} \frac{2}{\Gamma (j-1) \Gamma (b+j) \Gamma (m-j+1) \Gamma (a-j+m+2)}\sum _{i=1}^{m-j+1} \frac{\psi_0(i+j)}{i},
\end{eqnarray}
which is now the same form as~\eref{eq:a2d2next4b}. Inserting the result~\eref{eq:a2d2next4acancel1} into~\eref{eq:a2d2next4a} before adding up~\eref{eq:a2d2next4b}, we obtain
\begin{eqnarray}
\fl\sum _{j=1}^m \frac{2}{\Gamma (j-1) \Gamma (a+j) \Gamma (m-j+1) \Gamma (b-j+m+2)}\sum _{i=1}^{m-j+1} \frac{\psi _0(i+j)}{i}\nonumber\\
\fl+\sum _{j=1}^m \frac{2}{\Gamma (j-1) \Gamma (b+j) \Gamma (m-j+1) \Gamma (a-j+m+2)}\sum _{i=1}^{m-j+1} \frac{\psi _0(i+j)}{i}\nonumber\\
\fl=\sum _{j=1}^m \frac{1}{\Gamma (j-1) \Gamma (a+j) \Gamma (m-j+1) \Gamma (b-j+m+2)}\bigg(\left(\psi _0(m-j+2)+\psi _0(j)\right)\nonumber\\
\fl~~~\!~\times\left(\psi _0(m-j+2)+\psi _0(j)-2 \psi _0(1)\right) -\psi _1(m-j+2)-\psi _1(j)+2 \psi _1(1)\bigg),
\end{eqnarray}
which is simplified into single sums of the forms~\eref{basis1},~\eref{basis2},~\eref{basis3}, and
\begin{equation}
\sum _{j=1}^{m} \frac{\Gamma (a+b-j+2 m-1)}{ \Gamma (a-j+m)j}\psi_0(j) \label{basis5},
\end{equation}
by using lemma~\ref{lemma2}, lemma~\ref{lemma4}, and their derivatives with respect to $c$.
After inserting the simplified results of~\eref{eq:A2SS0},~\eref{eq:A2S2a}, and~\eref{eq:A2S2b} into~(\ref{eq:A2S2}), we observe complete cancellations of the single sums~\eref{basis1},~\eref{basis2},~\eref{basis3}, and~\eref{basis5}. The sum $\mathcal{A}_2(m-2,m)$ is simplified to
\begin{eqnarray}
\mathcal{A}_2(m-2,m)&=&\frac{4 \Gamma (a+m+1) \Gamma (b+m+1) }{\Gamma (m-1) \Gamma (a+b+m+1) (a+b+2 m-1)_3}\nonumber\\
&&\times\sum _{j=1}^{m} \frac{\psi_0 (a+b+j+m)}{j}+\mathrm{CF},
\end{eqnarray}
where the shorthand notation $\mathrm{CF}$, different in each use, denotes some closed-form terms omitted due to the length. Using the same approach, one is able to simplify~\eref{eq:A2S1} into a similar form, which completes the simplification of~$\mathcal{A}_2$ as per~\eref{eq:A2}.

Now inserting the resulting forms of $\mathcal{A}_1$ and $\mathcal{A}_2$ into~\eref{eq:IAex}, $\mathrm{I_{\mathcal{A}}}$ is finally obtained as
\begin{eqnarray}
\mathrm{I_{\mathcal{A}}}=f_\mathcal{A}(a,b)+f_\mathcal{A}(b,a),
\end{eqnarray}
where
\begin{eqnarray}\label{result:iaab}
\fl f_\mathcal{A}(a,b)&=&b_0 \sum _{k=1}^m \frac{\psi _0(a+b+k+m)}{a+k}-m \sum _{k=1}^m \frac{\psi _0(a+b+k+m)}{k}+b_1 \sum _{k=1}^m \frac{\psi _0(a+k)}{k}\nonumber\\
\fl &&+\frac{m}{2} \left(\psi _0^2(a+b+m)-\psi _1(a+b+m)\right)+b_2 \left(\psi _0(a+b+2 m)-\psi _0(a+b+m)\right)\nonumber\\
\fl &&\times\psi _0(a+b+2 m) +b_3 \psi _0(a+m) \psi _0(a+b+2 m)+b_4 \psi _0(a+b+2 m)\nonumber\\
\fl &&+b_0 \psi _0(a) \psi _0(a+b+m)+m \left(\psi _0(m)-\psi _0(1)\right) \psi _0(a+b+m)+b_5 \psi _0(a+m)\nonumber\\
\fl &&\times \left(2 \psi _0(a+b+m)+\psi _0(b+m)\right)+b_6 \psi _0(a+b+m)+\frac{b_1}{2} (2 \psi _0(a) \psi _0(a+m)\nonumber\\
\fl &&-2 \psi _0(a) \psi _0(m)-\psi _0^2(a)+2 \psi _0(1) \psi _0(a)+\psi _1(a))+b_7 \psi _0(a+m)+b_8 \psi _0(a)\nonumber\\
\fl &&+b_9.
\end{eqnarray}
The coefficients $b_i$ in~\eref{result:iaab} are summarized in~\eref{coeff_b0}--\eref{coeff_b9} in~appendix C.2.

By inserting~\eref{result:icab} and~\eref{result:iaab} into~\eref{eq:fpac2}, we obtain
\begin{eqnarray}\label{eq:fpprer}
\fl~~~~~~~\mathbb{E}\!\left[C\right]&=&\frac{2 m (a+m) (b+m) (a+b+m)}{(a+b+2 m-1)_3}\left(\sum _{k=1}^m \frac{\psi_0(a+k)}{k}+\sum _{k=1}^m \frac{\psi_0(b+k)}{k}\right.\nonumber\\
\fl~~~~~~~&&+\!\left.\sum _{k=1}^m \frac{\psi_0(a+b+k+m)}{a+k}+\sum _{k=1}^m \frac{\psi_0 (a+b+k+m)}{b+k}\right)+\mathrm{CF}.
\end{eqnarray}
The remaining task in obtaining~\eref{eq:prop1} is to represent the single summations
\begin{eqnarray}
\sum _{k=1}^m \frac{\psi_0(a+k)}{k}\label{eq:fpsum1}\\
\sum _{k=1}^m \frac{\psi_0(b+k)}{k}\label{eq:fpsum2}\\
\sum _{k=1}^m\frac{\psi_0(a+b+k+m)}{a+k}\label{eq:fpsum3}\\
\sum _{k=1}^m \frac{\psi_0 (a+b+k+m)}{b+k}\label{eq:fpsum4}
\end{eqnarray}
in~\eref{eq:fpprer} into~\eref{eq:sumpsiab} as reproduced below
\begin{equation}
\Phi _{c,d}=\frac{c! }{(c+d)!}\sum _{k=1}^c \frac{(c+d-k)!}{ (c-k)!}\frac{1}{k^2}, \qquad c,d\in \mathbb{Z^+}.
\end{equation}
By utilizing the identity~\eref{eq:B12}, the summations in~\eref{eq:fpsum1} and~\eref{eq:fpsum2} are respectively computed into the summations~$\Phi _{m,a}$ and $\Phi _{m,b}$ as
\begin{eqnarray}
\sum _{k=1}^m \frac{\psi_0(a+k)}{k}=\Phi _{m,a}+\mathrm{CF}\label{eq:fpsum11}\\
\sum _{k=1}^m \frac{\psi_0(b+k)}{k}=\Phi _{m,b}+\mathrm{CF}\label{eq:fpsum21}.
\end{eqnarray}
 To proceed with the summations~\eref{eq:fpsum3} and~\eref{eq:fpsum4}, we have to consider their combination
\begin{equation}
\sum _{k=1}^m\frac{\psi_0(a+b+k+m)}{a+k}+\sum _{k=1}^m \frac{\psi_0 (a+b+k+m)}{b+k}.
\end{equation}
We first rewrite~\eref{eq:fpsum3} as
\begin{eqnarray}\label{eq:fpsum31}
\!\!\!\!\!\!\!\!\!\!\!\!\!\!\sum _{k=1}^m\frac{\psi_0(a+b+k+m)}{a+k}=\sum _{k=1}^m \frac{\psi _0(b+k)}{a+k}+\sum _{k=1}^m \frac{1}{a+k}\sum _{l=0}^{a+m-1} \frac{1}{b+k+l},
\end{eqnarray}
where we have used the finite sum form of digamma function~\cite{Brychkov08}
\begin{equation}\label{eq:pl0}
\psi_{0}(l)=-\gamma+\sum_{k=1}^{l-1}\frac{1}{k}
\end{equation}
 to replace
\begin{equation}
\psi_0(a+b+k+m)
\end{equation}
by
\begin{equation}
\psi _0(b+k)+\sum _{l=0}^{a+m-1} \frac{1}{b+k+l}.
\end{equation}
We then change the order of summation of the double sum in~\eref{eq:fpsum31} to evaluate the sum over $k$ first, where the remaining sums are further evaluated by the identity~\eref{eq:B3}, leading to
\begin{eqnarray}\label{eq:fpsum32}
\fl \sum _{k=1}^m \frac{\psi_0(a+b+k+m)}{b+k}&=&\sum _{k=1}^{a+m-1} \frac{\psi _0(b+k+1)}{k}-\sum _{k=1}^{a+m-1} \frac{\psi _0(b+k+m+1)}{k}\nonumber\\
\fl&&+\frac{1}{2} \bigg( \!\left(2 \psi _0(1)-2 \psi _0(a+m)-\psi _0(b+m+1)-\psi _0(b+1)\right)\nonumber\\
\fl&&\times\left(\psi _0(b+1)-\psi _0(b+m+1)\right)-\psi _1(b+m+1)+\psi _1(b+1)\bigg).\nonumber\\
\end{eqnarray}
Similarly, one has~\eref{eq:fpsum4} manipulated to
\begin{eqnarray}\label{eq:fpsum41}
\fl \sum _{k=1}^m \frac{\psi_0(a+b+k+m)}{a+k}&=&\sum _{k=1}^{b+m-1} \frac{\psi _0(b+k+1)}{k}-\sum _{k=1}^{b+m-1} \frac{\psi _0(a+k+m+1)}{k}+\mathrm{CF}.
\end{eqnarray}
Here, we also need the result
\begin{eqnarray}\label{eq:fpsum42}
\fl& \sum _{k=1}^{a+m-1} \frac{\psi_0(b+k+m+1)}{k}+\sum _{k=1}^{b+m-1} \frac{\psi_0(a+k+m+1)}{k}\nonumber\\
\fl&=-\frac{1}{2}( \psi _1(a+m+1)+ \psi _1(b+m+1))-\frac{(a+b+2 m) \psi _0(a+b+2 m)+1}{(a+m) (b+m)}\nonumber\\
\fl&~~~\!~-\frac{1}{2} \left(2 \psi _0(1)-\psi _0(a+m+1)-\psi _0(b+m+1)\right) \left(\psi _0(a+m+1)+\psi _0(b+m+1)\right)\nonumber\\
\fl&~~~\!~+\psi _1(1),
\end{eqnarray}
which is obtained by evaluating the summation
\begin{equation}
\sum _{k=1}^{a+m-1} \frac{\psi _0(b+k+m+1)}{k}=\sum _{k=1}^{a+m-1} \frac{\psi _0(k)}{k}+\sum _{k=1}^{a+m-1} \frac{1}{k}\sum _{l=0}^{b+m} \frac{1}{ (k+l)}
\end{equation}
in the same manner as we have processed~\eref{eq:fpsum31}. Finally, by adding~\eref{eq:fpsum32} and~\eref{eq:fpsum41} before using~\eref{eq:fpsum42}, we obtain
\begin{eqnarray}
&&\sum _{k=1}^m \frac{\psi _0(a+b+k+m)}{a+k}+\sum _{k=1}^m \frac{\psi _0(a+b+k+m)}{b+k}\nonumber\\
&=&\sum _{k=1}^{b+m} \frac{\psi_0(a+k)}{k}+\sum _{k=1}^{a+m} \frac{\psi_0(b+k)}{k}+\mathrm{CF}.\\
&=&\Phi _{b+m,a}+\Phi _{a+m,b}+\mathrm{CF},\label{eq:fpsumf}
\end{eqnarray}
where the last equality~\eref{eq:fpsumf} is obtained by using the identity~\eref{eq:B12}. Inserting the results~\eref{eq:fpsum11}, \eref{eq:fpsum21}, and~\eref{eq:fpsumf} into~\eref{eq:fpprer}, we complete the proof of proposition~\ref{prop1}.

\subsection{Average capacity over fermionic Gaussian states without particle number constraint}\label{sec:2.2}
In this section, we compute the mean value of entanglement capacity~\eref{eq:capacity} over fermionic Gaussian states without particle number constraint~\eref{eq:fgap} in proving proposition~\ref{prop2}. The same as the previous section, we first discuss the computation that leads to the summation representation in~\sref{sec:2.2.1}. Simplification of the summations is performed in~\sref{sec:2.2.2}.

\subsubsection{Correlation functions and integral calculations}\label{sec:2.2.1}
For fermionic Gaussian states of arbitrary number of particles, by definition the average capacity is given by the integral
\begin{equation}\label{eq:EC20}
\mathbb{E}\!\left[C\right]=m\int_{0}^{1}u(x)g_1(x)\dd x,
\end{equation}
where $g_1(x_{1},\dots,x_{l})$ denotes the joint probability density of $l$ arbitrary eigenvalues. Similar to the previous case, the density $g_1(x_{1},\dots,x_{l})$ can be written in terms of the $l$-point correlation function as
\begin{equation}\label{eq:gl2}
g_{l}(x_{1},\dots,x_{l})=\frac{(m-l)!}{m!}\det\left(K\left(x_{i},x_{j}\right)\right)_{i,j=1}^{l},
\end{equation}
where
\begin{equation}\label{eq:corrk2}
K\left(x,y\right)=\sqrt{w(x)w(y)}\sum_{k=0}^{m-1}\frac{J_{2k}^{(a,a)}(x) J_{2k}^{(a,a)}(y)}{h_k}
\end{equation}
with the weight function being
\begin{equation}
w(x)=\left(\frac{1-x}{2}\right)^{a}  \left(\frac{1+x}{2}\right)^{a}.
\end{equation}
By rewriting the orthogonality relation~\eref{eq:orthogonality} as
\begin{eqnarray}
&\int_{0}^{1}\left(\frac{1-x}{2}\right)^{a}\left(\frac{1+x}{2}\right)^{a}J^{(a,a)}_{2k}(x)J^{(a,a)}_{2l}(x)\dd x\nonumber\\
&=\frac{\Gamma(2k+a+1)\Gamma(2k+a+1)}{(4k+2a+1)\Gamma(2k+1)\Gamma(2k+2a+1)}\delta_{kl}, \quad \Re(a)>-1,
\end{eqnarray}
we obtain the normalization constant $h_{k}$ of the polynomials $J^{(a,a)}_{2k}(x)$
\begin{equation}
h_{k}=\frac{\Gamma(2k+a+1)\Gamma(2k+a+1)}{(4k+2a+1)\Gamma(2k+1)\Gamma(2k+2a+1)}.
\end{equation}

By using \eref{eq:ux} and~\eref{eq:corrk2}, the computation of the average capacity~\eref{eq:EC20} boils down to computing two integrals
\begin{equation}\label{eq:EC2}
\mathbb{E}\!\left[C\right]=\mathrm{I_C}-\mathrm{I_A},
\end{equation}
where
\begin{eqnarray}
\mathrm{I_{C}}&=&\sum_{k=0}^{m-1}\frac{1}{h_{k}}\int_{-1}^{1}\left(\frac{1-x}{2}\right)^{a}\left(\frac{1+x}{2}\right)^{a+2}\ln^{2}\frac{1+x}{2}J_{2k}^{(a,a)}(x)^2\dd x\label{eq:apIc}\\
\mathrm{I_A}&=&\mathrm{A_1}+\mathrm{A_2}\label{eq:apA}
\end{eqnarray}
with
\begin{eqnarray}
\!\!\!\!\!\!\!\!\!\!\!\!\!\!\!\!\!\!\!\!\mathrm{A_{1}}&=&\sum_{k=0}^{m-1}\frac{1}{h_{k}}\int_{-1}^{1}\left(\frac{1-x}{2}\right)^{a}\left(\frac{1+x}{2}\right)^{a+2}\ln^{2}\frac{1+x}{2}J_{2k}^{(a,a)}(x)^2\dd x\label{eq:apA1}\\
\!\!\!\!\!\!\!\!\!\!\!\!\!\!\!\!\!\!\!\!\mathrm{A_{2}}&=&\sum_{k=0}^{m-1}\frac{1}{h_{k}}\int_{-1}^{1}\left(\frac{1-x}{2}\right)^{a+1}\left(\frac{1+x}{2}\right)^{a+1}\ln\frac{1-x}{2}\ln\frac{1+x}{2}J_{2k}^{(a,a)}(x)^2\dd x.\label{eq:apA2}
\end{eqnarray}
The integral in $\mathrm{I_{C}}$ is calculated by applying the identity~(\ref{eq:SIac2}), where we need to assign
\begin{equation}\label{eq:sp11}
a_1=b_1=a_2=b_2=a,\qquad k_1=k_2=2k,
\end{equation}
and take twice derivatives of $c$ before setting $c=a+1$. Under the same specialization~(\ref{eq:sp11}), the integral in $\mathrm{A_{1}}$ is calculated by taking twice derivatives of $c$ of the identity~(\ref{eq:SIac2}) before setting $c=a+1$, whereas the integral in $\mathrm{A_2}$ is calculated by taking derivatives of both $c$ and $d$ of the identity~(\ref{eq:SIdc2}) before setting $c=d=a+1$. After resolving the indeterminacy of gamma and polygamma functions by using~(\ref{eq:pgna1})--(\ref{eq:pgna3}), one arrives at the summation representations~(\ref{eq:AICS})--(\ref{eq:AA2S}) of the above integrals as listed in appendix A.2.

\subsubsection{Simplification of summations}\label{sec:2.2.2}
The remaining task in computing the mean value~\eref{eq:EC2} is to simplify the summation representations~\eref{eq:AICS}--\eref{eq:AA2S} of the integrals $\mathrm{I_{C}}$ and $\mathrm{I_{A}}$.

We first compute $\mathrm{I_{C}}$ by simplifying the summations in~\eref{eq:AICS}. Note that~\eref{eq:AICS} consists of two double summations. The first double summation is readily reduced to a single sum by evaluating the inner sum over $j$. The resulting single sum is further simplified by using the identities~\eref{eq:B1}--\eref{eq:B8} similarly to the simplification of~\eref{eq:IcabS}. Here, one will also need the results
\begin{eqnarray}
\psi_{0}(mk)=\ln m+\frac{1}{m}\sum_{i=0}^{m-1}\psi_{0}\!\left(k+\frac{i}{m}\right)\label{eq:m_poly0}, \qquad m\in\mathbb{Z^{+}}\label{eq:m_poly0}\\
\psi_{1}(mk)=\frac{1}{m^2}\sum _{i=0}^{m-1} \psi _1\!\left(\frac{i}{m}+k\right), \qquad m\in\mathbb{Z^{+}}\label{eq:m_poly0}
\end{eqnarray}
to evaluate the sums involving polygamma functions with even argument.
In~\eref{eq:AICS}, the second double sum is
\begin{eqnarray}\label{eq:AICS2}
\!\!\!\!\!\!\!\!\!\!\!\!\!\!\!\!\!\!\!\!\!\!\!&\sum _{k=1}^{m-1} 2 (2 a+4 k+1)\sum _{j=0}^{2 k-2} \frac{2 (j+1) (a+j+1)}{(2k-j-1)_2 (2 a+j+2 k+1)_2}\nonumber\\
\!\!\!\!\!\!\!\!\!\!\!\!\!\!\!\!\!\!\!\!\!\!\!&\times\left(\psi _0(a+j+2)-\psi _0(2 a+j+2 k+3)-\psi _0(2k-j-1)+\psi _0(j+2)\right).
\end{eqnarray}
By the partial fraction decomposition
\begin{eqnarray}
\!\!\!\!\!\!\!\!\!\!\!\!\!\!\!\!\!\!\!\!\!\!\!\!\!\frac{2 (j+1) (a+j+1)}{(2k-j-1)_2 (2 a+j+2 k+1)_2}\nonumber\\
\!\!\!\!\!\!\!\!\!\!\!\!\!\!\!\!\!\!\!\!\!\!\!\!\!=\frac{1}{2 a+4 k+1}\left(\frac{-2 a-2 k-1}{2 a+j+2 k+2}+\frac{2 (a+k)}{2 a+j+2 k+1}-\frac{2 k}{j-2 k+1}+\frac{2 k+1}{j-2 k}\right),
\end{eqnarray}
we rewrite~\eref{eq:AICS2} as the sum of the following five double summations~(\ref{eq:AICS21})--(\ref{eq:AICS25}),
\begin{eqnarray}
\!\!\!\!\!\!\!\!\!\!\!\!\!\!\!\!\!\!\!\!\!\!\!2\sum _{k=1}^{m-1} \sum _{j=0}^{2 k-2} \left(\frac{2a+2k+1}{2 a+j+2 k+2}-\frac{2 (a+k)}{2 a+j+2 k+1}\right) \psi_0(2 a+j+2 k+3)\label{eq:AICS21}\\
\!\!\!\!\!\!\!\!\!\!\!\!\!\!\!\!\!\!\!\!\!\!\!2\sum _{k=1}^{m-1} \sum _{j=0}^{2 k-2} \left(\frac{2 k}{j-2 k+1}-\frac{2 k+1}{j-2 k}\right) \psi_0(2k-j-1)\label{eq:AICS22}\\
\!\!\!\!\!\!\!\!\!\!\!\!\!\!\!\!\!\!\!\!\!\!\!2\sum _{k=1}^{m-1} \sum _{j=0}^{2 k-2} \left(\frac{2 k}{2k-j-1}-\frac{2 k+1}{2k-j}\right) \psi_0(2 a+j+2 k+3)\label{eq:AICS23}\\
\!\!\!\!\!\!\!\!\!\!\!\!\!\!\!\!\!\!\!\!\!\!\!2\sum _{k=1}^{m-1}\sum _{j=0}^{2 k-2} \left(\frac{2a+2k+1}{2 a+j+2 k+2}-\frac{2 (a+k)}{2 a+j+2 k+1}\right) \psi_0(2k-j-1)\label{eq:AICS24}\\
\!\!\!\!\!\!\!\!\!\!\!\!\!\!\!\!\!\!\!\!\!\!\!2\sum _{k=1}^{m-1} \sum _{j=0}^{2 k-2} \left(\frac{-2 a-2 k-1}{2 a+j+2 k+2}+\frac{2 (a+k)}{2 a+j+2 k+1}-\frac{2 k}{j-2 k+1}+\frac{2 k+1}{j-2 k}\right)\nonumber\\
\!\!\!\!\!\!\!\!\!\!\!\!\!\!\!\!\!\!\!\!\!\!\!\times (\psi _0(a+j+2)+\psi_0(j+2))\label{eq:AICS25}.
\end{eqnarray}
We now simplify each of the summations~\eref{eq:AICS21}--\eref{eq:AICS25} into single sums. Specifically, the summation~\eref{eq:AICS21} is simplified by using the identity~\eref{eq:B3} to evaluate the sum over $j$. The summation~\eref{eq:AICS22} is simplified similarly after shifting the index $j\to2k-2-j$. The summation~\eref{eq:AICS23} is simplified by using the identity~\eref{eq:B1} to evaluate the sum over $k$ after shifting the index $j\to2k-2-j$ and changing the summation order as
\begin{eqnarray}
2 \sum _{j=0}^{m-1} \sum _{k=j+1}^{m-1} \left(\frac{2 k}{2 j+1}-\frac{2 k+1}{2 j+2}\right) \psi _0(2 a-2 j+4 k+1)\nonumber\\
+2 \sum _{j=0}^{m-1} \sum _{k=j+1}^{m-1} \left(\frac{2 k}{2 j+2}-\frac{2 k+1}{2 j+3}\right) \psi _0(2 a-2 j+4 k),
\end{eqnarray}
where one has divided the summation over $j$ into even and odd ones. The remaining two sums \eref{eq:AICS24}--\eref{eq:AICS25} are simplified in a similar approach as~\eref{eq:AICS23}. For~\eref{eq:AICS24}, one needs to shift the index $j\to2k-2-j$ before changing the summation order to evaluate the sum over $k$. For~\eref{eq:AICS25}, one directly evaluates the sum over $k$ by changing the summation order.

Putting together the results of~\eref{eq:AICS21}--\eref{eq:AICS25}, the summation~\eref{eq:AICS} now consists of single sums, cf.~\eref{eq:IcabS}, which are further simplified by the identities~\eref{eq:B1}--\eref{eq:B8}. This leads to
\begin{eqnarray}\label{result:ic}
\fl \mathrm{I_C}&=&\sum _{k=1}^{m-1}\left(\left(-\frac{1}{4 k}-\frac{1}{4 k+2}\right) \psi _0(a+k)+\left(\frac{4 m-3}{4 k+2}+\frac{4 m+1}{4 k}\right) \psi _0(a+2 k)\right.\nonumber\\
\fl &&+\left(\frac{4 a+4 m-1}{2 a+4 k+2}+\frac{4 a+4 m-1}{2 a+4 k}-\frac{2 a}{2 k+1}+\frac{1-2 a}{2 k}+\frac{1}{2 (a+k)}\right) \psi _0(2 a+2 k)\nonumber\\
\fl&&+\left(\frac{2 a-1}{2 k}+\frac{2 a+1}{2 k+1}+\frac{-2 a-1}{2 a+2 k}+\frac{1-2 a}{2 a+2 k+1}\right) \psi _0(2 a+4 k)+\left(\frac{1}{4 k+2}\right.\nonumber\\
\fl&&-\!\left.\frac{1}{4 k}\right)\!\left.\psi _0(a+k+m)+\left(\frac{-2 a-2 m+1}{2 k}-\frac{2 a+2 m}{2 k+1}\right) \psi _0(2 a+2 k+2 m)\right)\nonumber\\
\fl&&+c_0 \psi _1(2 a+2 m)-\frac{1}{4} \psi _1(a+m)+c_1 \left(\psi _1(2 a)-\psi _0^2(2 a)\right)+c_2 \psi _1(a)\nonumber\\
\fl&&+c_3 \psi _0(2 a+4 m) \left(\psi _0(a+2 m)+\psi _0(2 a+2 m)-\psi _0(2 a+4 m)\right)-2 c_0 \psi _0(2 a+2 m)\nonumber\\
\fl&&\times\left(\psi _0(a)+\psi _0(2 m)-\psi _0(1)\right)-c_0 \psi _0^2(2 a+2 m)+ c_4 \psi _0^2(a+2 m)+c_5 \psi _0(a)\nonumber\\
\fl&&\times \psi _0(a+2 m)+\frac{1}{2} \left(\psi _0(a)+\psi _0(2 m)-\psi _0(1)\right) \psi _0(a+m)-\frac{1}{2} \psi _0(a) \psi _0(m)\nonumber\\
\fl&&+c_6 \psi _0(a) \psi _0(2 m)+c_7 \psi _0^2(a)+c_8 \psi _0(2 a+4 m)+c_9 \psi _0(2 a+2 m)+c_{10} \psi _0(a+2 m)\nonumber\\
\fl&&+c_{11} \psi _0(a+m)+c_{12} \psi _0(2 a)+c_{13} \psi _0(1) \psi _0(a)+c_{14} \psi _0(a)+c_{15} \left(\psi _0\left(\frac{a}{2}+m+\frac{1}{4}\right)\right.\nonumber\\
\fl&&-\!\left.\psi _0\left(\frac{a}{2}+\frac{1}{4}\right)\right)+c_{16} \psi _0\left(\frac{a}{2}+m\right)+c_{17} \left(\psi _0(m)-2 \psi _0(2 m)+\psi _0(1)\right)+c_{18} \psi _0\left(\frac{a}{2}\right)\nonumber\\
\fl&&-2 m,
\end{eqnarray}
where the coefficients $c_i$ are listed in~\eref{coeff_c0}--\eref{coeff_c18} in appendix C.3.

The simplification of~\eref{eq:AA1S} and \eref{eq:AA2S} in computing $\mathrm{I_A}$ is parallel to that of~\eref{eq:AICS} and~\eref{eq:A2S2}, respectively, where much of details are omitted here. However, we note that when first evaluating the inner summations over $i$ and $j$ in~\eref{eq:AA2S}, the resulting sum simply becomes
\begin{equation}
-r(k)\frac{2 \left(2 a^2+4 a k+a+4 k^2+2 k-1\right)}{(2 a+4 k-1) (2 a+4 k+3)}\sum _{j=1}^{2 k} \frac{\psi_0(2 a+j+2 k)}{j}+\mathrm{CF},
\end{equation}
where the term
\begin{equation}
r(k)=\frac{\Gamma (2 a+4 k+4)}{(2 a+4 k+1) \Gamma (2 k+1) \Gamma (2 a+2 k+1)}
\end{equation}
cancels completely with that in~\eref{eq:AA2S}. The remaining sums now only consist of rational functions and polygamma functions, which are readily simplifiable. Inserting the resulting forms of~\eref{eq:AA1S} and~\eref{eq:AA2S} into~\eref{eq:apA}, we obtain
\begin{eqnarray}\label{result:ia}
\fl\mathrm{I_A}&=&\sum _{k=1}^{m-1}\left(\left(-\frac{1}{2 (2 k+1)}-\frac{1}{4 k}\right) \psi _0(a+k)+\left(\frac{2 a m-2 a+6 m^2-6 m+1}{(2 k+1) (2 a+4 m-1)}\right.\right.\nonumber\\
\fl&&+\!\!\left.\frac{1}{4 (a+k)}+\frac{4 a m+2 a+12 m^2-1}{4 k (2 a+4 m-1)}\right) \psi _0(a+2 k)+\left(\frac{1-2 a}{2 k}+\frac{1}{2 (a+k)}-\frac{2 a}{2 k+1}\right.\nonumber\\
\fl&&+\!\left.\frac{2 \left(2 a^2+5 a m-a+3 m^2-m\right)}{2 a+4 m-1}\left(\frac{1}{a+2 k+1}+\frac{1}{a+2 k}\right) \right) \psi _0(2 a+2 k)\nonumber\\
\fl&&+\left(\frac{2 a-1}{2 k}+\frac{-2 a-1}{2 (a+k)}+\frac{2 a+1}{2 k+1}+\frac{1-2 a}{2 a+2 k+1}\right) \psi _0(2 a+4 k)+\left(\frac{1}{2 (2 k+1)}\right.\nonumber\\
\fl&&-\!\left.\frac{1}{4 k}\right) \!\left.\psi _0(a+k+m)+\left(\frac{-2 a-2 m+1}{2 k}-\frac{2 a+2 m}{2 k+1}\right) \psi _0(2 a+2 k+2 m)\right)\nonumber\\
\fl&&+d_0 \psi _1(2 a+2 m)+d_1 \left(\psi _1(2 a)-\psi _0^2(2 a)\right)+d_2 \psi _1(a)-\frac{1}{4} \psi _1(a+m)\nonumber\\
\fl&&+d_3 \left(\psi _0(a+2 m)+\psi _0(2 a+2 m)-\psi _0(2 a+4 m)\right) \psi _0(2 a+4 m)+d_0\psi _0(2 a+2 m) \nonumber\\
\fl&&\times\left(-\psi _0(2 a+2 m)-2 \psi _0(2 m)+2 \psi _0(1)\right) +d_4 \psi _0(a+2 m) \psi _0(2 a+2 m)\nonumber\\
\fl&&+d_5 \psi _0^2(a+2 m)+d_6 \psi _0(a) \psi _0(2 a+2 m)+a \psi _0(a) \left(\psi _0(a)-2 \psi _0(a+2 m)\right)\nonumber\\
\fl&&+\frac{1}{4} \left(\psi _0(a)+2 \psi _0(2 m)-2 \psi _0(1)\right) \psi _0(a+m)+d_7 \psi _0(a) \psi _0(2 m)-\frac{1}{4} \psi _0(a) \psi _0(m)\nonumber\\
\fl&&+d_8 \psi _0(2 a+4 m)+d_9 \psi _0(2 a+2 m)+d_{10} \psi _0(a+2 m)+d_{11} \psi _0(a+m)\nonumber\\
\fl&&+d_{12} \left(\psi _0(m)-2 \psi _0(2 m)+\psi _0(1)\right)+d_{13} \psi _0(2 a)+d_{14} \psi _0(1) \psi _0(a)+d_{15} \psi _0(a)+d_{16}\nonumber\\
\fl&& \times \left(\psi _0\left(\frac{a}{2}+m+\frac{1}{4}\right)-\psi _0\left(\frac{a}{2}+\frac{1}{4}\right)\right)+d_{17} \left(\psi _0\left(\frac{a}{2}\right)-\psi _0\left(\frac{a}{2}+m\right)\right)-m,
\end{eqnarray}
where the coefficients $d_i$ are listed in~\eref{coeff_d0}--\eref{coeff_d17} in appendix C.4.

Inserting the results~\eref{result:ic} and~\eref{result:ia} into~\eref{eq:EC2}, the mean capacity becomes
\begin{eqnarray}\label{result:EC2}
\fl \mathbb{E}\!\left[C\right]&=&\frac{m (a+m)}{2 a+4 m-1}\sum _{k=1}^{m-1} \frac{\psi _0(a+2 k)}{k}-\frac{1}{4}\sum _{k=1}^{m-1} \frac{\psi _0(a+2 k)}{a+k}+\frac{(2 m-1) (2 a+2 m-1)}{2 (2 a+4 m-1)}\nonumber\\
\fl&&\times\sum _{k=1}^{m-1}\left( \frac{\psi _0(a+2 k+1)}{2 k+1}+ \frac{\psi _0(2 a+2 k)}{a+2 k}+\frac{\psi _0(2 a+2 k+1)}{a+2 k+1}\right)+\mathrm{CF},
\end{eqnarray}
where we recall that the shorthand notation $\mathrm{CF}$ denotes the closed-form terms omitted. In the above result~\eref{result:EC2}, we rewrite the single summations
\begin{equation}
\sum _{k=1}^{m-1} \frac{\psi _0(a+2 k+1)}{2 k+1}\label{eq:ec2s1}
\end{equation}
and
\begin{equation}
\sum _{k=1}^{m-1} \frac{\psi _0(2 a+2 k+1)}{a+2 k+1}\label{eq:ec2s2}
\end{equation}
as
\begin{equation}
\sum _{k=1}^{m-1} \frac{\psi _0(a+2 k+1)}{2 k+1}=\sum _{k=2}^{2 m} \frac{\psi _0(a+k)}{k}-\frac{1}{2} \sum _{k=1}^m \frac{\psi _0(a+2 k)}{k} \label{eq:2m1}
\end{equation}
and
\begin{eqnarray}
\sum _{k=1}^{m-1} \frac{\psi _0(a+2 k+1)}{2 k+1}&=\sum _{k=2}^{2 m} \frac{\psi _0(2 a+k)}{a+k}-\sum _{k=1}^m \frac{\psi _0(2 a+2 k)}{a+2 k}\\
&=\sum _{k=1}^{a+2 m} \frac{\psi _0(a+k)}{k}-\sum _{k=1}^m \frac{\psi _0(2 a+2 k)}{a+2 k}+\mathrm{CF},\label{eq:a2m1}
\end{eqnarray}
respectively.
Here, the equality~\eref{eq:a2m1} is obtained by shifting the summation index as
\begin{equation}
\!\!\!\!\!\!\!\!\sum _{k=2}^{2 m} \frac{\psi _0(2 a+k)}{a+k}=\sum _{k=2+a}^{2 m+a} \frac{\psi _0( a+k)}{k}=\sum _{k=1}^{2 m+a} \frac{\psi _0( a+k)}{k}-\sum _{k=1}^{a+1} \frac{\psi _0( a+k)}{k},
\end{equation}
before evaluating the last sum by the identity~\eref{eq:B5}. Moreover, for the summation
\begin{equation}
\sum _{k=1}^{m-1} \frac{\psi _0(a+2 k)}{k}\label{eq:ec2s3},
\end{equation}
we have
\begin{eqnarray}
\!\!\!\!\!\!\!\!\!\!\!\!\!\!\!\!\!\!\!\!\sum _{k=1}^{m-1} \frac{\psi _0(a+2 k)}{k}= \sum _{k=1}^{m-1} \left(\frac{\psi _0(a+k+m)}{k}+\frac{\psi _0(a+k)}{k}+\frac{\psi _0(a+2 k)}{a+k}\right)+\mathrm{CF},\label{eq:a2k1}
\end{eqnarray}
which is obtained by the fact that
\begin{equation}
\sum _{k=1}^{m-1} \frac{\psi _0(a+2 k)}{a+k}=\sum _{k=1}^{m-1} \sum _{l=0}^{k-1} \frac{1}{(a+k) (a+k+l)}+\sum _{k=1}^{m-1} \frac{\psi_0(a+k)}{a+k}
\end{equation}
similarly to the identity~\eref{eq:fpsum31}. By substituting in~\eref{result:EC2} the sums~\eref{eq:ec2s1},~\eref{eq:ec2s2},~and~\eref{eq:ec2s3} with their equivalent forms~\eref{eq:2m1},~\eref{eq:a2m1}, and~\eref{eq:a2k1}, respectively, we arrive at
\begin{eqnarray}\label{result:EC22}
\!\!\!\!\!\!\!\!\!\!\!\!\!\!\!\!\!\!\!\mathbb{E}\!\left[C\right]&=&\frac{(2 m-1) (2 a+2 m-1)}{4 a+8 m-2}\left(\sum _{k=1}^{2 m-1} \frac{\psi_0(a+k)}{k}+\sum _{k=1}^{2m+a-1} \frac{\psi_0(a+k)}{k}\right)\nonumber\\
\fl &&+\frac{1}{4} \left(\sum _{k=1}^{m-1} \frac{\psi_0(a+k+m)}{k}+\sum _{k=1}^{m-1} \frac{\psi_0(a+k)}{k}\right)+\mathrm{CF}.
\end{eqnarray}
Finally, replacing the single sums in~\eref{result:EC22} by the short-hand notation $\Phi_{c,d}$ defined in~(\ref{eq:sumpsiab}) the claimed result~\eref{eq:prop2} is obtained. This completes the proof of proposition~\ref{prop2}.

\subsection{Asymptotic capacity}\label{sec:2.3}
In this section, we compute the limiting average capacity in corollary~\ref{corollary1}. This is a rather straightforward task starting from the exact formula of average capacity. Specifically, the limiting values in~\eref{eq:limit1} are obtained by computing the limits of the exact capacity~\eref{eq:prop1} and~\eref{eq:prop2} in the regime~\eref{regime1}. To this end, the following asymptotic results are needed. The first one is the limiting behavior of polygamma functions
\begin{eqnarray}
\psi_0(x)=&\ln (x)-\frac{1}{2 x}-\sum_{l=1}^{\infty}\frac{B_{2l}}{2lx^{2l}},\qquad x\to\infty \label{eq:limpl0},\\
\psi_{1}(x)=&\frac{1+2x}{2x^{2}}+\sum_{l=1}^{\infty}\frac{B_{2l}}{x^{2l+1}},\qquad x\to\infty \label{eq:limpl1},
\end{eqnarray}
where $B_k$ is the $k$-th Bernoulli number~\cite{AS72}. The second one is the fact that in the asymptotic regime
\begin{equation}
c\to\infty,\qquad \mathrm{with~a~fixed}~d,
\end{equation}
one has
\begin{equation}
\Phi _{c,d}\longrightarrow\psi_1(1)=\frac{\pi^2}{6}.
\end{equation}

For the exact capacity formula~\eref{eq:prop1} of fermionic Gaussian states with fixed particle number~\eref{eq:fgfp}, we now have in the limit~\eref{regime1},
\begin{eqnarray}
\frac{\alpha_0}{m}=\frac{1}{8}+o\left(\frac{1}{m}\right)\\
\frac{\alpha_1}{m}=o\left(\frac{1}{m}\right)\\
\frac{\alpha_2}{m}=o\left(\frac{1}{m}\right)\\
\frac{\alpha_3}{m}=-\frac{1}{2}+o\left(\frac{1}{m}\right),
\end{eqnarray}
and
\begin{eqnarray}
\psi _1(a+b+m+1)+\psi _1(a+m+1)=o\left(\frac{1}{m}\right)\\
\psi _0(a+m+1)-\psi _0(a+b+m+1)=o\left(\frac{1}{m}\right)\\
\psi _0(a+m+1)=o(\ln m),
\end{eqnarray}
where we recall $a=n-p$ and $b=p-m$.
Consequently, we obtain
\begin{eqnarray}
\mathbb{E}\!\left[C\right]&=&2 \left(\frac{1}{8}+o\left(\frac{1}{m}\right)\right) \left(\frac{\pi ^2}{2}+o\left(\frac{1}{m}\right)\right)\nonumber\\&&+2o\left(\frac{1}{m}\right) o(\ln m)-1+o\left(\frac{1}{m}\right),
\end{eqnarray}
where, by using the fact that
\begin{equation}
\displaystyle{\lim_{m \to \infty}}\frac{\ln m}{m}=0,
\end{equation}
one arrives at the claimed asymptotic result
\begin{equation}
\mathbb{E}\!\left[C\right]\stackrel{\eref{regime1}}{\longrightarrow}\frac{\pi^2}{8}-1.
\end{equation}

For the exact capacity~\eref{eq:prop2} of fermionic Gaussian states with arbitrary particle number~\eref{eq:fgap}, similarly we have in the limit~\eref{regime1},
\begin{eqnarray}
\psi _1(m+n)=o\left(\frac{1}{m}\right)\\
\psi _1(n)=o\left(\frac{1}{m}\right)\\
\psi _0(2 n)-\psi _0(m+n)=o\left(\frac{1}{m}\right)\\
\psi _0(m+n)-\psi _0(n)=\ln 2 +o\left(\frac{1}{m}\right)\\
\psi _0(m+n)-\psi _0(n-m)=-\psi_0(n-m)+\ln2+o(\ln m).
\end{eqnarray}
As a result, we have
\begin{eqnarray}
\mathbb{E}\!\left[C\right]&=&\frac{1}{3} \pi ^2 \left(o\left(\frac{1}{m}\right)+\frac{1}{2}\right)+\left(o\left(\frac{1}{m}\right)+\frac{1}{4}\right) \left(o\left(\frac{1}{m}\right)-\frac{\pi ^2}{6}\right)\nonumber\\
&&+o\left(\frac{1}{m}\right) o(\ln m)+o\left(\frac{1}{m}\right)-1,
\end{eqnarray}
which leads to the claimed result
\begin{equation}
\mathbb{E}\!\left[C\right]\stackrel{\eref{regime1}}{\longrightarrow}\frac{\pi^2}{8}-1.
\end{equation}
This completes the proof of corollary~\ref{corollary1}.

\section{Conclusion}
In this work, we derived the exact and asymptotic average capacity formulas of fermionic Gaussian states with and without particle number constraints. The derivation of the results relies on tools from random matrix theory and, more importantly, recent progress in simplifying finite summations involving special functions. The obtained analytical formulas provide insights into the statistical behavior of entanglement as measured by entanglement capacity. Future works include computing higher-order statistics, such as the variance, of entanglement capacity.

\section*{Acknowledgments}
The work of Lu Wei is supported in part by the U.S. National Science Foundation ($\#$2150486).

\appendix

\section{Summation representations of integrals}\label{App_SE}
In this appendix, we list the summation representations of the integrals $\mathrm{I_{\mathcal{C}}}^{(a,b)}$, $\mathcal{A}_1^{(a,b)}$, $\mathcal{A}_2$ in~\eref{eq:Iccd}--\eref{eq:A2} and $\mathrm{I_C}$, $\mathrm{A_1}$, $\mathrm{A_2}$ in~\eref{eq:apIc}, \eref{eq:apA1},~\eref{eq:apA2} in the computation of average entanglement capacity in section~\ref{sec:2}.

\subsection{Summation representations of integrals $\mathrm{I_{\mathcal{C}}}^{(a,b)}$, $\mathcal{A}_1^{(a,b)}$, $\mathcal{A}_2(m-1,m-1)$, and $\mathcal{A}_2(m-2,m)$}\label{app:SE1}
\begin{eqnarray} \label{eq:IcabS}
\fl \mathrm{I_{\mathcal{C}}}^{(a,b)}&=&\frac{2 m (b+m)}{a+b+2 m}\sum _{i=1}^{m-2} \frac{i}{(m-i-1)_2}\left(\psi _0(b+i+1)-\psi _0(m-i-1)+\psi _0(i+1)\right.\nonumber\\
\fl&&-\!\left.\psi _0(a+b+i+m+1)\right)-\frac{(a+m) (a+b+m)}{a+b+2 m}\sum _{i=1}^{m-1} \frac{2 i}{(a+b+i+m)_2}\nonumber\\
\fl &&\times\left(\psi _0(b+i+1)-\psi _0(m-i)+\psi _0(i+1)-\psi _0(a+b+i+m+2)\right)\nonumber\\
\fl &&+\frac{m (b+m)}{a+b+2 m} \sum _{i=m}^{m+1} \frac{(i-1) (-1)^{i+m-1}}{\Gamma (i-m+1) \Gamma (m-i+2)}\bigg(\psi _1(b+i)-\psi _1(i-m+1)\nonumber\\
\fl&&+\psi _1(i)-\psi _1(a+b+i+m)+(\psi _0(b+i)-\psi _0(i-m+1)+\psi _0(i)\nonumber\\
\fl &&-\psi _0(a+b+i+m)){}^2\bigg)
\end{eqnarray}

\begin{eqnarray}\label{eq:A1abS}
\fl \mathcal{A}_1^{(a,b)}&=&-\frac{2m(b+m)}{a+b+2 m}\sum _{i=1}^{m-3} \frac{(b+i+1) (i)_2}{(m-i-2)_3 (a+b+i+m+1)}(\psi_0(b+i+2)+\psi_0(i+2)\nonumber\\
\fl&&-\psi_0(a+b+i+m+2)-\psi_0(m-i-2))+\frac{2 (a+m)( a+b+m)}{a+b+2 m}\nonumber\\
\fl&&\times\sum _{i=1}^{m-2} \frac{(b+i+1) (i)_2}{(m-i-1) (a+b+i+m)_3}(\psi_0(b+i+2)-\psi_0(a+b+i+m+3)\nonumber\\
\fl&&-\psi_0(m-i-1)+\psi_0(i+2))-\frac{m (b+m)}{a+b+2 m}\sum _{i=m-3}^{m-1} \frac{(b+i+2) (-1)^{i+m} (i+1)_2}{\Gamma (m-i) \Gamma (i-m+4) }\nonumber\\
\fl&&\times\frac{1}{a+b+i+m+2}\big(\psi _1(b+i+3)-\psi _1(i-m+4)-\psi _1(a+b+i+m+3)\nonumber\\
\fl&&+(\psi_0(i+3)-\psi_0(a+b+i+m+3)-\psi_0(i-m+4)+\psi_0(b+i+3))^2\big)\nonumber\\
\fl&&+\psi _1(i+3)-\frac{(a + m)(a+b+m) (b+m) (m-1)_2}{(a+b+2 m) (a+b+2 m-1)_3}\bigg(\!-\psi _1(a+b+2 m+2)\nonumber\\
\fl&&+\psi _1(b+m+1)+\psi _1(m+1)-\psi _1(1)+\psi_0^2(1)+(\psi _0(b+m+1)+\psi _0(m+1)\nonumber\\
\fl&&-\psi _0(a+b+2 m+2))(\psi _0(b+m+1)+\psi _0(m+1)-\psi _0(a+b+2 m+2)\nonumber\\
\fl&&-2 \psi _0(1))\bigg)
\end{eqnarray}

\begin{eqnarray}\label{eq:A2S1}
\fl&\!\!\!\!\!\mathcal{A}_2(m-1,m-1)\nonumber\\
\fl=&\frac{2 \Gamma (a+m+1) \Gamma (b+m+1)}{\Gamma (a+b+2 m+2)}\Bigg(\sum _{i=1}^m \frac{i (m-i+1)(-1)^{i}}{\Gamma (a+i+1) \Gamma (b-i+m+2)}\sum _{j=i-2}^i (-1)^{j}\nonumber\\
\fl&\times\frac{ \Gamma (a+i-j+m) \Gamma (b-i+j+m+2)}{\Gamma (j+1) \Gamma (i-j+1) \Gamma (j-i+3) \Gamma (m-j)}\bigg(\psi _1(a+b+2 m+2)\nonumber\\
\fl&+\left(\psi _0(a+i-j+m)-\psi _0(i-j+1)+\psi _0(i+1)-\psi _0(a+b+2 m+2)\right) \nonumber\\
\fl&\times(\psi _0(a+b+2 m+2)-\psi _0(b-i+j+m+2)+\psi _0(j-i+3)\nonumber\\
\fl&-\psi _0(m-i+2))\bigg)+\sum _{i=1}^{m-2} \frac{i (m-i+1)}{\Gamma (b+i+1) \Gamma (a-i+m+2)}\!\sum _{j=1}^{m-i-1}\frac{\Gamma (b-j+m)}{\Gamma (m-i-j)}\nonumber\\
\fl&\times \frac{\Gamma (a+j+m+2)}{(j)_3 \Gamma (i+j+1)}(\psi _0(a+j+m+2)+\psi _0(m-i+2)-\psi _0(j+3)\nonumber\\
\fl&-\psi _0(a+b+2 m+2))+\sum _{i=1}^{m-2} \frac{i (m-i+1)}{\Gamma (a+i+1) \Gamma (b-i+m+2)}\nonumber\\
\fl&\times\sum _{j=1}^{m-i-1} \frac{\Gamma (a-j+m) \Gamma (b+j+m+2)}{(j)_3 \Gamma (i+j+1) \Gamma (m-i-j)}(\psi _0(b+j+m+2)+\psi _0(m-i+2)\nonumber\\
\fl&-\psi _0(j+3)-\psi _0(a+b+2 m+2))\Bigg)
\end{eqnarray}

\begin{eqnarray}\label{eq:A2S2}
\fl &\!\!\!\!\!\mathcal{A}_2(m-2,m)\nonumber\\
\fl=&\frac{2 \Gamma (a+m) \Gamma (b+m)}{\Gamma (a+b+2 m+2)}\Bigg(\sum _{i=1}^{m-1} \frac{\Gamma (a+m+1) \Gamma (b+m+1)}{\Gamma (i) \Gamma (a+i+1) \Gamma (m-i) \Gamma (b-i+m+1)}\nonumber\\
\fl&\times\!\bigg((\psi _0(a+b+2 m+2)-\psi _0(a+m+1)-\psi _0(i+1)+\psi _0(1)) (\psi _0(m-i+1)\nonumber\\
\fl&-\psi _0(a+b+2 m+2)+\psi _0(b+m+1)-\psi _0(1))+\psi _1(a+b+2 m+2)\bigg)\nonumber\\
\fl&+\sum _{i=1}^{m-1} \frac{i (m-i)}{\Gamma (b+i+1) \Gamma (a-i+m+1)}\sum _{j=1}^{m-i} \frac{\Gamma (a+j+m+1) \Gamma (b-j+m+1)}{j \Gamma (i+j+1) \Gamma (m-i-j+1)}\nonumber\\
\fl &\times\left(\psi _0(a+j+m+1)-\psi _0(a+b+2 m+2)+\psi _0(m-i+1)-\psi _0(j+1)\right)\nonumber\\
\fl&+\sum _{i=1}^{m-1} \frac{i (m-i)}{\Gamma (a+i+1) \Gamma (b-i+m+1)}\sum _{j=1}^{m-i} \frac{\Gamma (a-j+m+1) \Gamma (b+j+m+1)}{j \Gamma (i+j+1) \Gamma (m-i-j+1)}\nonumber\\
\fl &\times(\psi _0(b+j+m+1)-\psi _0(a+b+2 m+2)+\psi _0(m-i+1)-\psi _0(j+1))\Bigg)
\end{eqnarray}

\subsection{Summation representations of integrals $\mathrm{I_C}$, $\mathrm{A_1}$, and $\mathrm{A_2}$}\label{app:SE2}
\begin{eqnarray}\label{eq:AICS}
\fl \mathrm{I_C}=&\!\left(\psi _0(a+2)-\psi _0(2 a+3)\right)^2+\psi _1(a+2)-\psi _1(2 a+3)+\sum _{k=1}^{m-1} 2 (2 a+4 k+1)\nonumber\\
\fl &\times\left(\rule{0cm}{0.75cm}\sum _{j=2 k-1}^{2 k} \frac{(-1)^j (j+1) (a+j+1)}{(2 a+j+2 k+1)_2}\bigg((\psi _0(j+2)-\psi _0(2 a+j+2 k+3)\right.\nonumber\\
\fl &+\psi _0(a+j+2)-\psi _0(j-2 k+2))^2+\psi _1(a+j+2)-\psi _1(2 a+j+2 k+3)\nonumber\\
\fl &+\psi _1(j+2)-\psi _1(j-2 k+2)\bigg)+\sum _{j=0}^{2 k-2} \frac{2 (j+1) (a+j+1)}{(2k-j-1)_2 (2 a+j+2 k+1)_2}\nonumber\\
\fl & \times\!\left.\left(\psi _0(a+j+2)-\psi _0(2 a+j+2 k+3)-\psi _0(2k-j-1)+\psi _0(j+2)\right)\rule{0cm}{0.75cm}\right)
\end{eqnarray}

\begin{eqnarray}\label{eq:AA1S}
\fl \mathrm{A_1}=&\sum _{k=0}^{m-1}2(2 a+4 k+1)\left(\rule{0cm}{0.75cm}\sum _{j=2 k-2}^{2 k}\frac{(-1)^j (j+1)_2 (a+j+1)_2}{\Gamma (2k-j+1) \Gamma (j-2 k+3) (2 a+j+2 k+1)_3}\right.\nonumber\\
\fl &\times\bigg((\psi_0 (a+j+3)-\psi_0 (2 a+j+2 k+4)-\psi_0 (j-2 k+3)+\psi_0 (j+3))^2\nonumber\\
\fl &-\psi _1(2 a+j+2 k+4)+\psi _1(a+j+3)-\psi _1(j-2 k+3)+\psi _1(j+3)\bigg)\nonumber\\
\fl & +\sum _{j=0}^{2 k-3} \frac{2 (j+1)_2 (a+j+1)_2}{(2k-j-2)_3 (2 a+j+2 k+1)_3}(\psi_0 (2 a+j+2 k+4)-\psi_0 (a+j+3)\nonumber\\
\fl &+\!\left.\psi_0 (2k-j-2)-\psi_0 (j+3))\rule{0cm}{0.75cm}\right)
\end{eqnarray}

\begin{eqnarray}\label{eq:AA2S}
\fl \mathrm{A_2}=&\sum _{k=0}^{m-1} \frac{(2 a+4 k+1) \Gamma (2 k+1) \Gamma (2 a+2 k+1)}{\Gamma (2 a+4 k+4)}\left(\rule{0cm}{0.75cm}\sum _{i=0}^{2 k}\frac{2 (i+1) (2k-i+1) }{\Gamma (i+1)\Gamma (a+i+1)}\right.\nonumber\\
\fl &\times\frac{\Gamma^{2}(a+2 k+2)}{\Gamma(2k-i+1)\Gamma(a+2k-i+1)}((\psi_0 (a+2 k+2)-\psi_0 (2 a+4 k+4)-\psi_0 (2)\nonumber\\
\fl &+\psi_0 (2k-i+2)) (\psi_0 (a+2 k+2)-\psi_0 (2 a+4 k+4)+\psi_0 (i+2)-\psi_0 (2))\nonumber\\
\fl &-\psi _1(2 a+4 k+4))-\sum _{j=0}^{2 k} \frac{(j+1) \Gamma (a+2 k+1) \Gamma (a+2 k+3)}{\Gamma (j) \Gamma (a+j+1) \Gamma (2k-j+1)\Gamma (a-j+2 k+1)}\nonumber\\
\fl &\times ((\psi_0 (a+2 k+1)-\psi_0 (2 a+4 k+4)+\psi_0 (2k-j+2)-\psi_0 (1)) (\psi_0 (a+2 k+3)\nonumber\\
\fl &-\psi_0 (2 a+4 k+4)+\psi_0 (j+2)-\psi_0 (3))-\psi _1(2 a+4 k+4))-\sum _{j=0}^{2 k} \frac{(2k-j+1)}{\Gamma (j+1)}\nonumber\\
\fl &\times\frac{ \Gamma (a+2 k+1) \Gamma (a+2 k+3)}{ \Gamma (a+j+1) \Gamma (2 k-j) \Gamma (2 k-j+a+1)}((\psi_0 (a+2 k+3)-\psi_0 (2 a+4 k+4)\nonumber\\
\fl & +\psi_0 (2k-j+2)-\psi_0 (3)) (\psi_0 (a+2 k+1)-\psi_0 (2 a+4 k+4)+\psi_0 (j+2)\nonumber\\
\fl &-\psi_0 (1))-\psi _1(2 a+4 k+4))+4 \sum _{j=0}^{2 k} \frac{\Gamma (a-j+2 k) \Gamma (a+j+2 k+4)}{(j+1)_3}\nonumber\\
\fl&\times\sum _{i=0}^{2k-j-2}\frac{ (2 k-i-j-1)(i+j+3) }{\Gamma (i+1)  \Gamma (2k-i+1) \Gamma (a+i+j+3) \Gamma (a-i-j+2 k-1)}\nonumber\\
\fl &\times\!\left.(\psi_0 (a+j+2 k+4)-\psi_0 (2 a+4 k+4)+\psi_0 (i+j+4)-\psi_0 (j+4))\rule{0cm}{0.75cm}\right).
\end{eqnarray}

\section{List of summation identities}\label{App_SI}
In this appendix, we list the finite sum identities useful in simplifying the summations in appendix A. Proofs to these identities can be found, for example, in~\cite{Wei17,Wei20,Wei20BH,HWC21,Milgram, HW22, HW23}. Here, it is sufficient to assume $a,b\ge0, a\neq b$ in identities~\eref{eq:B1}--\eref{eq:B3},~\eref{eq:B6}--\eref{eq:B7}, $a>m$ in~\eref{eq:B8}, and $a,b\geq0$, $n>m$ in~\eref{eq:B9}--\eref{eq:B14}.

\begin{eqnarray}\label{eq:B1}
\fl\sum_{i=1}^{m}\psi_{0}(i+a)=(m+a)\psi_{0}(m+a+1)-a\psi_{0}(a+1)-m
\end{eqnarray}
\begin{equation}\label{eq:B2}
\fl\sum_{i=1}^{m}\psi_{1}(i+a)=(m+a)\psi_{1}(m+a+1)-a\psi_{1}(a+1)+\psi_{0}(m+a+1)-\psi_{0}(a+1)
\end{equation}
\begin{eqnarray}\label{eq:B3}
\fl\sum_{i=1}^{m}\frac{\psi_{0}(i+a)}{i+a}=\frac{1}{2}\left(\psi_{1}(m+a+1)-\psi_{1}(a+1)+\psi_{0}^{2}(m+a+1)-\psi_{0}^{2}(a+1)\right)
\end{eqnarray}
\begin{eqnarray}\label{eq:B4}
\fl\sum_{i=1}^{m}\frac{\psi_{0}(m+1-i)}{i}=\psi_{0}^{2}(m+1)-\psi_{0}(1)\psi_{0}(m+1)+\psi_{1}(m+1) -\psi_{1}(1)
\end{eqnarray}
\begin{eqnarray}\label{eq:B5}
\fl\sum _{i=1}^m \frac{\psi_0 (m+1+i)}{i}=\psi _0^2(m+1)-\psi _0(1) \psi _0(m+1)-\frac{1}{2} \psi _1(m+1)+\frac{\psi _1(1)}{2}
\end{eqnarray}
\begin{eqnarray}\label{eq:B6}
\fl\sum_{i=1}^{m}\psi_{0}(i+a)\psi_{0}(i+b)&=&(b-a) \sum _{i=1}^{m-1} \frac{\psi_0(a+i)}{b+i}+(m+a) \psi_0(m+a) \psi_0(m+b)-a\times\nonumber\\
&& \psi_0(a+1) \psi_0(b+1)-(m+a-1) \psi_0(m+a)+a \psi_0(a+1)-\nonumber\\
&&(m+b) \psi_0(m+b)+(b+1) \psi_0(b+1)+2 m-2
\end{eqnarray}
\begin{eqnarray}\label{eq:B7}
\fl \sum_{i=1}^{m}\frac{\psi_{0}(i+b)}{i+a}=&-\sum_{i=1}^{m}\frac{\psi_{0}(i+a)}{i+b}+\psi_{0}(m+a+1)\psi_{0}(m+b+1)-\psi_{0}(a+1)\times\nonumber\\
&\psi_{0}(b+1)+\frac{1}{a-b}(\psi_{0}(m+a+1)-\psi_{0}(m+b+1)-\psi_{0}(a+1)+\nonumber\\
&\psi_{0}(b+1))
\end{eqnarray}
\begin{eqnarray}\label{eq:B8}
\fl \sum_{i=1}^{m}\frac{\psi_{0}(a+1-i)}{i}=&-\sum_{i=1}^{m}\frac{\psi_{0}(i+a-m)}{i}+(\psi_{0}(a-m)+\psi_0(a+1))(\psi_{0}(m+1)-\nonumber\\
&\psi_{0}(1))+\frac{1}{2}\left((\psi_{0}(a-m)-\psi_{0}(a+1))^2+\psi_{1}(a+1)-\psi_{1}(a-m)\right) \nonumber\\
\end{eqnarray}

\begin{eqnarray}\label{eq:B9}
\fl\sum _{i=1}^m \frac{\psi_0(a+b+i)}{i}=&\sum _{i=1}^m \frac{\psi_0(b+i)}{i}-\sum _{i=1}^a \frac{\psi_0(b+i+m)}{b+i-1}+\frac{1}{2} \bigg(\psi _1(b)+(\psi_0(a+b)-\psi_0(b))\nonumber\\
\fl&\times (\psi_0(a+b)+\psi_0(b)+2 (\psi_0(m+1)-\psi_0(1)))-\psi _1(a+b)\bigg)
\end{eqnarray}

\begin{eqnarray}\label{eq:B10}
\fl\sum _{i=1}^{m} \frac{(n-i)!}{ (m-i)!}=\frac{n!}{(m-1)! (n-m+1)}
\end{eqnarray}

\begin{eqnarray}\label{eq:B11}
\fl\sum _{i=1}^{m} \frac{(n-i)!}{ (m-i)!i}=\frac{n! }{m!}(\psi _{0}(n+1)-\psi _{0}(n-m+1))
\end{eqnarray}

\begin{eqnarray}\label{eq:B12}
\fl\sum _{i=1}^{m} \frac{(n-i)!}{ (m-i)!i^2}=&\frac{n! }{m!} \Bigg(\sum _{i=1}^m\frac{\psi_0(i+n-m)}{i}+\frac{1}{2}\big (\psi _1(n-m+1)-\psi _1(n+1)-\psi _0^2(n+1)\nonumber\\
\fl&+\psi _0^2(n-m+1)\big)+\psi _0(n-m)(\psi _0(n+1)-\psi _0(n-m+1)\nonumber\\
\fl&-\psi _0(m+1)+\psi _0(1))\Bigg)
\end{eqnarray}

\begin{eqnarray}\label{eq:B13}
\fl\sum _{i=1}^m \frac{(n-i)!}{(m+a-i)!}=\frac{1}{n-m-a+1}\left(\frac{n!}{(a+m-1)!}-\frac{(n-m)!}{(a-1)!}\right)
\end{eqnarray}

\begin{eqnarray}\label{eq:B14}
\fl&\sum _{i=1}^m \frac{(n-i)! }{(m+a-i)!}\psi _0(m+a-i+1)\nonumber\\
\fl&=\frac{1}{1-a-m+n}\left(\frac{n! }{(a+m-1)!}\left(\psi _0(a+m)-\frac{1}{1-a-m+n}\right)-\frac{(n-m)! }{(a-1)!}\right.\nonumber\\
\fl&~~~\!~\times\!\left.\left(\psi _0(a)-\frac{1}{1-a-m+n}\right)\right)
\end{eqnarray}

\section{Coefficients of results in section~\ref{sec:2}}\label{app:coeff}
In this appendix, we list the coefficients in the results~\eref{result:icab},~\eref{result:iaab},~\eref{result:ic}, and~\eref{result:ia}.

\subsection{Coefficients in~\eref{result:icab}}\label{app:coeff_a}
\begin{eqnarray}
\fl a_0&=&\frac{2 (a+m) (a+b+m)}{a+b+2 m}\label{coeff_a0}\\
\fl a_1&=&\frac{2 m (b+m)}{a+b+2 m}\\
\fl a_2&=&\frac{2 \left(a^2+b (a+2 m)+2 a m+2 m^2\right)}{a+b+2 m}\\
\fl a_3&=&-\frac{2}{(b+m) (a+b+2 m)^2} \bigg(b^2 \left(a^2+8 a m+a+10 m^2\right)+2 a^2 m (m+2)+a^3+b^4\nonumber\\
\fl&&+b^3 (2 a+5 m)+b \left(a^2 (3 m+2)+6 a m (2 m+1)+2 m^2 (5 m+1)\right)+6 a m^2 (m+1)\nonumber\\
\fl&&+2 m^3 (2 m+1)\bigg)\\
\fl a_4&=&\frac{2 }{b (a+b+2 m)^2}\bigg(b^2 \left(a^2+a (5 m+2)+m (5 m+3)\right)+b (m+2) \left(a^2+3 a m+2 m^2\right)\nonumber\\
\fl&&+b^3 (2 a+4 m+1)+(a+m)^2 (a+2 m)+b^4\bigg)\\
\fl a_5&=&\frac{2 (b+m) \left(a^2+b (2 a+3 m)+3 a m+b^2+m (2 m-1)\right)}{(a+b+2 m)^2}\\
\fl a_6&=&-\frac{2 b (a+b+2 m+1)}{a+b+2 m}\\
\fl a_7&=&-\frac{2 m (a+m) \left(a^2+2 a b+4 a m+a+b^2+4 b m+b+4 m^2+2 m+1\right)}{(a+b+2 m)^3}\label{coeff_a7}
\end{eqnarray}

\subsection{Coefficients in~\eref{result:iaab}}\label{app:coeff_b}
\begin{eqnarray}
\fl b_0&=&\frac{2 (b+m) }{(a+b+2 m-1)_3}\left(a^2 (3 b+4 m)+a^3+a \left(3 b^2+9 b m+6 m^2-1\right)+5 b^2 m+b^3\right.\nonumber\\
\fl &&+\!\left.7 b m^2-b+3 m^3-m\right) \label{coeff_b0}\\
\fl b_1&=&\frac{2 m (a+m) \left(a^2+a (b+3 m)+2 b m+3 m^2-1\right)}{(a+b+2 m-1)_3}\\
\fl b_2&=&2 (a+m)\\
\fl b_3&=&-\frac{2 \left(a (b+2 m)+b^2+2 b m+2 m^2\right)}{a+b+2 m}\\
\fl b_4&=&\frac{m-m^2}{2 (a+b+2 m-1)}+\frac{m^2+m}{2 (a+b+2 m+1)}-\frac{2b}{a+m}-\frac{2m}{a+b+2 m}-2a-2m\\
\fl b_5&=&\frac{m (a+m) (b+m) (a+b+m)}{(a+b+2 m-1)_3}\\
\fl b_6&=&\frac{1}{8 a}\left(\frac{-2 a^2 b^2+a^4+4 a^2+b^4+4 b^2}{a+b+2 m}-\frac{1}{2} (a-b-1) (a-b+1) (a+b-1) (a+b+1) \right.\nonumber\\
\fl &&\times\!\left.\left(\frac{1}{a+b+2 m+1}+\frac{1}{a+b+2 m-1}\right)+16 a b+8 a m+7 a+11 b+6 m\right)\\
\fl b_7&=&\frac{b}{a+b+2 m}+\frac{(a-b) (a+b)}{2 (a+b+2 m)^2}+\frac{1}{4} (5 a+b-1)+m+\frac{1}{8} ((a-b-1) (a-b+1)\nonumber\\
\fl &&\times (a+b+1)) \left(\frac{1}{a+b+2 m+1}-\frac{1}{a+b+2 m-1}\right)\\
\fl b_8&=&-\frac{a \left(a^2+3 a (b+2 m+1)+2 b^2+b (6 m+3)+6 m^2+6 m+2\right)}{(a+b+2 m) (a+b+2 m+1)}\\
\fl b_9&=&-\frac{m}{(a+b+2 m)^2}-\frac{m}{2 (a+b+2 m+1)}-\frac{m}{2}\label{coeff_b9}
\end{eqnarray}

\subsection{Coefficients in~\eref{result:ic}} \label{app:coeff_c}
\begin{eqnarray}
\fl c_0&=&-\frac{1}{2} (2 a+2 m-1)\label{coeff_c0}\\
\fl c_1&=&\frac{1}{2} (2 a-1)\\
\fl c_2&=&\frac{1}{4} (4 m+1)\\
\fl c_3&=&-2 a-4 m+1\\
\fl c_4&=&\frac{1}{4} (4 a+4 m-1)\\
\fl c_5&=&-2 a\\
\fl c_6&=&1-2 m\\
\fl c_7&=&\frac{1}{4} (4 a-1)\\
\fl c_8&=&-2 (a+2 m)\\
\fl c_9&=&-\frac{-12 a^3-6 a^2+4 a+1}{4 a^3+6 a^2+2 a}\\
\fl c_{10}&=&\frac{a^2 (4 m-1)+4 a^3+a-1}{2 (a-1) a}\\
\fl c_{11}&=&\frac{a^4 (12-8 m)+a^3 (3-8 m)+a^2 (2 m-13)-12 a^5+2 a m+1}{2 \left(4 a^5-5 a^3+a\right)}\\
\fl c_{12}&=&\frac{a (8 (a+1) m+2 a+3)+2 m}{a (a+1) (2 a+1)}\\
\fl c_{13}&=&\frac{1}{2} (4 m-1)\\
\fl c_{14}&=&\frac{-4 m-3}{4 (a-1)}-2 a-\frac{3}{4 (a+1)}-\frac{1}{2 a-1}-\frac{1}{2 a+1}-\frac{1}{a}+\frac{1}{2} (4 m-3)\\
\fl c_{15}&=&-\frac{4 a^2+1}{4 a^2-1}\\
\fl c_{16}&=&-\frac{a^2 (4 m+1)+4 a^3+a (4 m-1)-1}{2 a \left(a^2-1\right)}\\
\fl c_{17}&=&\frac{m}{1-a}\\
\fl c_{18}&=&\frac{8 m+3}{4 (a-1)}+\frac{1}{2 a}-\frac{3}{4 (a+1)}+2\label{coeff_c18}
\end{eqnarray}

\subsection{Coefficients in~\eref{result:ia}}\label{app:coeff_d}
\begin{eqnarray}
\fl d_0&=&-\frac{1}{2} (2 a+2 m-1)\label{coeff_d0}\\
\fl d_1&=&a-\frac{1}{2}\\
\fl d_2&=&\frac{4 a^2-1}{16 (2 a+4 m-1)}-\frac{a}{8}+\frac{3 m}{4}+\frac{3}{16}\\
\fl d_3&=&-(2 a+4 m-1)\\
\fl d_4&=&\frac{(2 m-1) (2 a+2 m-1)}{4 a+8 m-2}\\
\fl d_5&=&\frac{1}{4} (4 a+4 m-1)\\
\fl d_6&=&\frac{(2 a+2 m-1) (4 a+6 m-1)}{2 (2 a+4 m-1)}\\
\fl d_7&=&\frac{1-4 a^2}{8 (2 a+4 m-1)}+\frac{a}{4}+\frac{1}{8} (5-12 m)\\
\fl d_8&=&-2 (a+2 m)\\
\fl d_9&=&-\frac{-12 a^3-6 a^2+4 a+1}{4 a^3+6 a^2+2 a}\\
\fl d_{10}&=&\frac{4 a^2 \left(4 m^2-2 m+1\right)+a^3 (20 m-3)+6 a^4-4 a (m-1)^2-4 m+1}{2 (a-1) a (2 a+4 m-1)}\\
\fl d_{11}&=&\frac{88 a^5-76 a^4-10 a^3+83 a^2-3 a-4}{16 (1-a) a (a+1) (2 a-1) (2 a+1)}-\frac{(2 a-1) (2 a+1)}{16 (a-1) (2 a+4 m-1)}-\frac{3 m}{4 (a-1)}\\
\fl d_{12}&=&-\frac{a (4 m+2)+12 m^2-1}{4 (a-1) (2 a+4 m-1)}\\
\fl d_{13}&=&\frac{8 a^3+4 a^2-2 a-1}{8 a (a+1) (2 a+4 m-1)}+\frac{3 (4 m-1)}{8 (a+1)}+\frac{3 (4 m+1)}{8 a}+\frac{4}{2 a+1}-\frac{1}{2}\\
\fl d_{14}&=&\frac{1-4 a^2}{-16 a-32 m+8}-\frac{a}{4}+\frac{1}{8} (12 m-3)\\
\fl d_{15}&=&\frac{8 a^3-12 a^2-2 a+3}{16 (a-1) (2 a+4 m-1)}-\frac{36 a^3+24 a^2-5 a-2}{4 a (a+1) (2 a-1) (2 a+1)}-\frac{12 m+13}{16 (a-1)}-\frac{7 a}{4}\nonumber\\
\fl&&+\frac{1}{4} (6 m-4)\\
\fl d_{16}&=&\frac{1+4 a^2}{1-4 a^2}\\
\fl d_{17}&=&\frac{4 a^2-1}{8 (a-1) (2 a+4 m-1)}+\frac{12 m+7}{8 (a-1)}+\frac{1}{4 a}-\frac{3}{4 (a+1)}+\frac{7}{4}\label{coeff_d17}.
\end{eqnarray}


\section*{References}
\begin{thebibliography}{99}

\bibitem{Page93}
Page D N 1993 Average entropy of a subsystem {\it \PRL}{\bf 71} 1291-4

\bibitem{Foong94}
Foong S K and Kanno S 1994 Proof of Page's conjecture on the average entropy of a subsystem {\it \PRL}{\bf 72} 1148-51

\bibitem{Ruiz95}
S\'{a}nchez-Ruiz J 1995 Simple proof of Page's conjecture on the average entropy of a subsystem {\it \PR E} {\bf 52} 5653-5

\bibitem{HLW06}
Hayden P, Leung D and Winter 2006 A Aspects of generic entanglement. {\it Commun. Math. Phys.} {\bf 265} 95-117.

\bibitem{VPO16}
Vivo P, Pato M P and Oshanin G 2016 Random pure states: Quantifying bipartite entanglement beyond the linear statistics {\it \PR E} {\bf 93} 052106

\bibitem{Wei17}
Wei L 2017 Proof of Vivo-Pato-Oshanin's conjecture on the fluctuation of von Neumann entropy {\it \PR E} {\bf 96} 022106

\bibitem{SK2019}
Sarkar A and Kumar S 2019 Bures-Hall ensemble: spectral densities and average entropies {\it \jpa}{\bf 52} 295203

\bibitem{Wei20}
Wei L 2020 Skewness of von Neumann entanglement entropy {\it \jpa}{\bf 53} 075302

\bibitem{Wei20BHA}
Wei L 2020 Proof of Sarkar-Kumar conjectures on average entanglement entropies over the Bures-Hall ensemble {\it \jpa}{\bf 53} 235203

\bibitem{Wei20BH}
Wei L 2020 Exact variance of von Neumann entanglement entropy over the Bures-Hall measure {\it \PR E} {\bf 102} 062128

\bibitem{HWC21}
Huang Y, Wei L and Collaku B 2021  Kurtosis of von Neumann entanglement entropy {\it \jpa}{\bf 54} 504003

\bibitem{Lubkin78}
Lubkin E 1978 Entropy of an n-system from its correlation with a k-reservoir {\it \JMP}{\bf 19} 1028

\bibitem{Sommers04}
Sommers H-J and \.{Z}yczkowski K 2004 Statistical properties of random density matrices {\it \JPA}{\bf 37} 35

\bibitem{Osipov10}
Osipov V, Sommers H-J and \.{Z}yczkowski K 2010 Random Bures mixed states and the distribution of their purity
 {\it \jpa}{\bf 43} 055302

\bibitem{Giraud07}
Giraud O 2007 Distribution of bipartite entanglement for random pure states {\it \jpa}{\bf 40} 2793

\bibitem{LW21}
Li S-H and Wei L 2021 Moments of quantum purity and biorthogonal polynomial recurrence {\it \jpa}{\bf 54} 445204

\bibitem{MML02}
Malacarne L C, Mendes R S and Lenzi E K 2002 Average entropy of a subsystem from its average Tsallis entropy {\it \PR E} {\bf 65} 046131

\bibitem{Wei19T}
Wei L 2019 On the exact variance of Tsallis entanglement entropy in a random pure state {\it Entropy} {\bf 21} 539

\bibitem{Borot12}
Borot G and Nadal C 2012 Purity distribution for generalized random Bures mixed states {\it \jpa}{\bf 45} 075209

\bibitem{BHK21}
Bianchi E, Hackl L and Kieburg M 2021 The Page curve for fermionic Gaussian states {\it \PR B} {\bf 103} L241118

\bibitem{HW22}
Huang Y and Wei L 2022 Second-order statistics of fermionic Gaussian states {\it \jpa}{\bf 55} 105201

\bibitem{BHKRV22}
Bianchi E, Hackl L, Kieburg M, Rigol M and Vidmar L 2022 Volume-law entanglement entropy of typical pure quantum states {\it PRX Quantum} {\bf 3} 030201

\bibitem{HW23}
Huang Y and Wei L 2022  Entropy fluctuation formulas of fermionic Gaussian states (arXiv:2211.16709)

\bibitem{YQ10}
Yao H and Qi X-L 2010 Entanglement entropy and entanglement spectrum of the Kitaev model {\it \PRL}{\bf 105} 080501

\bibitem{Nandy}
Nandy P 2021 Capacity of entanglement in local operators {\it \JHEP}19

\bibitem{ADKT23}
Arias R, Di Giulio G, Keski-Vakkuri E and Tonni E 2023 Probing RG flows, symmetry resolution and quench dynamics through the capacity of entanglement (arXiv:2301.02117)

\bibitem{Boer19}
de Boer J, J\"{a}rvel\"{a} J and Keski-Vakkuri E 2019 Aspects of capacity of entanglement {\it \PR D} {\bf 99} 066012

\bibitem{OKUYAMA21}
Okuyama K 2021 Capacity of entanglement in random pure state {\it \PL B} {\bf 820} 136600

\bibitem{Wei23}
Wei L 2023  Average capacity of quantum entanglement {\it \jpa}{\bf 56} 015302

\bibitem{BNP21}
Bhattacharjee B, Nandy P and Pathak T 2021 Eigenstate capacity and Page curve in fermionic Gaussian states {\it \PR B} {\bf 104} 214306

\bibitem{ST21}
Surace J and Tagliacozzo L 2021 Fermionic Gaussian states: An introduction to numerical approaches (arXiv:2111.08343)

\bibitem{AS72}
Abramowitz M and Stegun I A 1972 {\it Handbook of Mathematical Functions with Formulas, Graphs, and Mathematical Tables} (New York: Dover)

\bibitem{BZ06}
Bengtsson I and \.{Z}yczkowski K 2017 {\it Geometry of Quantum States: An Introduction to Quantum Entanglement} 2nd edn (Cambridge: Cambridge University Press)

\bibitem{KFI19}
Kieburg M, Forrester P and Ipsen J R 2019 Multiplicative convolution of real asymmetric and real anti-symmetric matrices {\it Adv. Pure Appl. Math.} {\bf 10} 467

\bibitem{LRV20}
{\L}yd\.{z}ba P, Rigol M and Vidmar L 2020 Eigenstate entanglement entropy in random quadratic Hamiltonians {\it \PRL} {\bf 125} 180604

\bibitem{LRV21}
{\L}yd\.{z}ba P, Rigol M and Vidmar L 2021 Entanglement in many-body eigenstates of quantum-chaotic quadratic Hamiltonians. {\it \PR B} {\bf 103} 104206



\bibitem{Mehta}
Mehta M L 2004 {\it Random Matrices} 3rd edn (Amsterdam: Elsevier)

\bibitem{Forrester}
Forrester P 2010 {\it Log-gases and Random Matrices} (Princeton: Princeton University Press)

\bibitem{BP21}
Bernard D and Piroli L 2021 Entanglement distribution in the quantum symmetric simple exclusion process {\it \PR E} {\bf 104} 014146


\bibitem{Szego}
Szeg{\H{o}} G 1975 {\it Orthogonal Polynomials} (Provindence: American Mathematical Society)

\bibitem{Luke}
Luke Y L 1969 {\it The Special Functions and Their Approximations Vol. 1} (Academic Press, New York)

\bibitem{Brychkov08}
Brychkov Y A 2008 {\it Handbook of Special Functions: Derivatives, Integrals, Series and Other Formulas} (Boca Raton: CRC Press)

\bibitem{Milgram}
Milgram M 2017 On some sums of digamma and polygamma functions (arXiv:0406338v3)

\end{thebibliography}
\end{document}